\documentclass[12pt,draftcls, onecolumn,dvips]{IEEEtran}

\usepackage{times}
\usepackage{amsmath,dsfont}
\usepackage{amssymb,amsthm}
\usepackage{epsfig,verbatim}
\usepackage{subfigure}
\usepackage{setspace}
\usepackage{breqn}

\newcommand{\mA}{\mathcal{A}}

\newcommand{\E}{\mathds{E}}

\long\def\symbolfootnote[#1]#2{\begingroup\def\thefootnote{\fnsymbol{footnote}}\footnote[#1]{#2}\endgroup}

 \IEEEoverridecommandlockouts
\title{Frequency-Shift Filtering for OFDM Signal Recovery in Narrowband Power Line Communications
\thanks{
Nir Shlezinger and Ron Dabora are with the Department of Electrical and Computer Engineering, Ben-Gurion University, Israel; Email: {\tt nirshl@post.bgu.ac.il}, {\tt ron@ee.bgu.ac.il}.
This work was supported by the Ministry of Economy of Israel through the Israeli Smart Grid Consortium.
}
}

\vspace{-0.5cm}

\author{Nir Shlezinger and Ron Dabora,~\IEEEmembership{Member,~IEEE}
\vspace{-1.0cm}
}

\vspace{-0.5cm}

\begin{document}

\maketitle
\pagestyle{plain}
\thispagestyle{plain}

\vspace{-1.0cm}

\begin{abstract}
\vspace{-0.5cm}
		Power line communications (PLC) has been drawing considerable interest in recent years due to the growing interest in smart grid implementation. Specifically, network control and grid applications are allocated the frequency band of $0-500$ kHz, commonly referred to as the narrowband PLC channel. This frequency band is characterized by strong periodic noise which results in low signal to noise ratio (SNR). In this work we propose a receiver which uses frequency shift filtering to exploit the cyclostationary properties of \emph{both} the narrowband power line noise, as well as the information signal, digitally modulated using orthogonal frequency division multiplexing. An adaptive implementation for the proposed receiver is presented as well. The proposed receiver is compared to existing receivers via analysis and simulation. The results show that the receiver proposed in this work obtains a substantial performance gain over previously proposed receivers, without requiring any coordination with the transmitter.
\end{abstract}
\vspace{-0.75cm}
\section{Introduction}
\vspace{-0.25cm}
In recent years the power supply network is changing its role from a network used solely for energy distribution into a dual-purpose network which simultaneously supports both communications as well as power distribution.
Generally speaking, power line communication (PLC) can be classified into two types, according to its frequency band~\cite{Pavlidou:03}. The first type is communication which utilizes low to medium frequencies (up to 500 kHz), this type is referred to as {\em narrowband PLC}. Such systems are used for applications of automation and control, including power management, smart homes, and automatic meter reading systems. The second type of communication  uses the frequency band of 2MHz to 100MHz and possibly beyond~\cite{OMEGA:08}. This type is referred to as {\em broadband PLC}, and is used for high-speed data communications including fast internet access and implementation of small LANs.

Power line communications differs considerably in topology and physical properties from conventional wired communication media such as twisted pair, coaxial, or fiber-optic cables. One of the major differences is the characteristics of interference and noise which are much more dominant in PLC than in other media~\cite{OMEGA:08}.
Furthermore, the statistical properties of the additive power line noise are considerably different than the conventionally used additive white Gaussian noise (AWGN) model~\cite{Pavlidou:03},~\cite{Hooijen:98},~\cite{Zimmermann:02a}.
%
As detailed in~\cite{Hooijen:98} and further developed in~\cite{Zimmermann:02b} and~\cite{Katayama:06}, power line noise can be modeled as a superposition of several noise sources:
\begin{itemize}
\item {Colored background noise: This is a low power noise whose power spectral density (PSD) decreases with frequency. This noise results mainly from the summation of harmonics of the mains cycle.}
\item {Narrowband noise: This noise consists of sinusoidal signals with modulated amplitudes, and is due to the operation of electrical appliances connected to the network (e.g., television related disturbances at high harmonics of the horizontal retrace frequency~\cite{Hooijen:98}).}
\item {Impulsive noise: This noise consists of impulses of varying duration, and is generated mostly by power supplies in electrical appliances. There are three types of impulsive noise:
(1) {\em Periodic impulsive noise synchronous with the AC cycle}: This noise is caused by the rectifier diodes used in power supplies which operate synchronously with the mains cycle. The impulses are of short duration (several microseconds) and their power decreases with frequency.
(2) {\em Periodic impulsive noise asynchronous with the AC cycle}: This noise is generated by switched-mode power supplies and AC/DC power converters, and has a cycle frequency that can vary between 50 to 200 kHz.
(3) {\em Non-periodic impulsive noise}: This noise is caused by switching transients. and has no periodic properties.
}
\end{itemize}
The impulsive noise components are the most harmful for broadband PLC~\cite{Zimmermann:02b}.
For narrowband PLC, the colored background noise, the narrowband noise and the periodic impulsive noise synchronous with the AC frequency are the dominant noise components~\cite{Katayama:06},~\cite{Joshi:08}. Due to the relatively long symbol duration in narrowband PLC transmission, the periodic properties of the noise cannot be ignored. One of the common models for the narrowband PLC noise is based on the work of Middleton in~\cite{Middleton:77}, which models the noise probability density function (PDF) as a sum of Gaussian PDFs of different variances, allowing to express several classes of impulsive noise by a simple function. This results in a non-Gaussian noise model. The drawback of the Middleton model is that it does not include time-domain features.  This issue has been addressed in the work of Katayama et al.~\cite{Katayama:06}, which proposed a time-domain cyclostationary noise model for the narrowband PLC noise. A recent work~\cite{Nassar:12} suggests an alternative cyclostationary noise model, obtained by applying a linear periodic time varying (LPTV) system to a stationary Gaussian stochastic process. Both works model the noise as an additive cyclostationary Gaussian noise (ACGN) with a period of half the mains period.
Finally, we note that due to the relatively high power of the power line noise, narrowband PLC typically operates at very low SNRs~\cite{Pavlidou:03}.

In the past several years, orthogonal frequency division multiplexing (OFDM) has been adopted for PLC schemes in order to achieve high bandwidth efficiency. OFDM is particularly suitable for coping with the frequency selectivity of the power-line channel~\cite{Amirshahi:06}.
An OFDM PLC modem structure was proposed as early as 1999~\cite{Deinzer:99}. The technology has  been adopted by recent narrowband PLC standardization efforts, IEEE P1901.2~\cite{IEEE:13} and ITU-T G.9903/4~\cite{G3:12, PRIME:12}.
However, due to the severe PLC noise, narrowband OFDM PLC is still limited to very low rates. Clearly, the implementation of smart grids poses significant data transfer requirements.
Therefore, the design of algorithms for handling the severe noise conditions in narrowband PLC is essential for the widespread implementation and realization of smart grids.

\vspace{-0.5cm}
\subsection *{Main Contributions and Organization}
\vspace{-0.25cm}
In the present paper we propose a receiver algorithm, based on the time-averaged mean squared error (TA-MSE) criterion, for recovery of OFDM signals received over the narrowband PLC channel. The receiver uses a cyclic version of the Wiener filter, called frequency shift (FRESH) filter, for exploiting the cyclostationary properties of the received signal. Specifically, we present the \emph{first} receiver designed for PLC which takes advantage of the cyclostationary properties of \emph{both} the {\em noise}, as well as the {\em information signal}. The novel idea is to utilize the cyclostationary properties of the noise to achieve noise reduction. This is generally {\em not possible} when the noise is {\em not cyclostationary} (e.g., for AWGN). The processing is specifically designed for low SNRs which characterize narrowband PLC. By exploiting the cyclostationary properties of both the OFDM signal and the noise, we achieve a substantial SNR gain compared to the receiver proposed in~\cite{Chen:11}, which used only the cyclostationarity of the OFDM signal. This is achieved without changing anything at the transmitter, thereby maintaining the spectral efficiency of the OFDM signal. The method proposed for cyclostationary noise reduction can be applied in both coded and uncoded narrowband PLC systems. We also show that the method is beneficial irrespective of the particular model of the cyclostationary noise, as long as cyclostationarity is maintained.

It is well known that Wiener filtering suffers from scaling of the signal at the output of the filter~\cite[Ch. 12.7]{Kay:93} which degrades the BER performance. In this work we derive analytically the scaling factor, which helps in estimating the actual SNR gain. We also discuss the application of the proposed receiver to channels with inter-symbol interference (ISI). Finally, we consider an adaptive implementation of the proposed algorithm and analyze the relationship between BER and TA-MSE. This indicates to the strength of the error correcting codes needed to obtain improved performance at different SNRs. Our results show the benefits of noise cancellation based on the noise properties, which should be considered in the design of future receivers for PLC.

The rest of this paper is organized as follows: in Section \ref{sec:Preliminaries} we briefly recall the relevant aspects of cyclostationarity to be used in this work. The cyclostationary properties of both the narrowband PLC noise and the OFDM signal are presented and the frequency shift filter is reviewed. In Section \ref{sec:Model}, the novel receiver algorithm is developed and its theoretical performance characteristics are obtained; and in Section \ref{sec:Adaptive} an adaptive implementation of the new receiver is discussed. We note that the design steps and algorithms used in the present work hold for both models~\cite{Katayama:06} and~\cite{Nassar:12} and the adaptive filter we propose works optimally for both models. In Section \ref{sec:Simulations} simulation results are presented together with a discussion. Lastly, conclusions are provided in Section \ref{sec:Conclusions}.

\vspace{-0.25cm}
\section{Preliminaries}
\label{sec:Preliminaries}
\vspace{-0.25cm}
\vspace{-0.25cm}
\subsection{Notations}
\label{sec:Notations}
\vspace{-0.25cm}
In the following we denote vectors with lower-case boldface letters, e.g., $\textbf{x},\textbf{y}$; the $i$-th element of a vector $\textbf{x}$ is denoted with $(\textbf{x})_i$. Matrices are denoted with upper-case boldface letters, e.g., $\textbf{X},\textbf{Y}$; the  element at the $i$-th row and the $j$-th column of a matrix $\textbf{X}$ is denoted with $(\textbf{X})_{i,j}$.
$(\cdot)^H$ denotes the Hermitian conjugate, $(\cdot)^T$ denotes the transpose, and $(\cdot)^*$ denotes the complex conjugate.
We use ${\mathop{\rm Re}\nolimits} \left\{x\right\}$ and ${\mathop{\rm Im}\nolimits} \left\{x\right\}$  to denote the real and imaginary parts of the complex number $x$ respectively, and $\mathds{Z}$ to denote the set of integers.
Lastly, $\delta[\cdot] $ denotes the Kronecker delta function, $\E\big\{ \cdot \}$ denotes the stochastic expectation, $\left\langle \cdot \right\rangle$ denotes the time-average operator, and ${\bf{1}}_A[\cdot]$ denotes the indicator function of a set $A$.

\vspace{-0.4cm}
\subsection{Cyclostationary Signals}
\label{sec:Cyclostationarity}
\vspace{-0.25cm}
A complex-valued discrete-time process $x[n]$ is said to be \textit{wide sense second order cyclostationary} (referred to henceforth as cyclostationary) if both its mean value and autocorrelation function are periodic with some integer period, say $N_0$, that is
$\E\big\{x[n]\} = \E\big\{x[n + N_0]\}$, and $c_{xx}(n,l) = \E\big\{x[n + l]x^*[n]\} = c_{xx}(n + N_0,l)$.
As $c_{xx}(n,l)$ is periodic in the variable $n$, it has a Fourier series expansion, whose coefficients, referred to as \textit{cyclic autocorrelation function}, are $c_{xx}^{\alpha _k}(l) = \frac {1} {N_0}\displaystyle\sum\limits_{n=0}^{N_0-1} {c_{xx}(n,l)e^{-j 2\pi \alpha _k n} }$, where
$\alpha _k = \frac {k} {N_0}, k=0,1,...,N_0-1,$ are referred to as the {\em cyclic frequencies}.

\vspace{-0.4cm}
\subsection{Cyclostationarity of OFDM Signals}
\label{subsec:cyclo_OFDM}
\vspace{-0.25cm}
Let $N_{data}$ denote the number of sub-carriers in an OFDM symbol and $N_{CP}$ denote the length of the cyclic prefix (CP). Then, $N_{sym} = N_{data} + N_{CP}$ is the length of an OFDM symbol in time samples. Let $a_{m,k}$ denote the data symbol at the $k$-th sub-carrier of the $m$-th OFDM symbol, and $q[n]$ be a real valued pulse shaping function of length $N_{sym}$ defined by $q[n] = {\bf{1}}_{\left\{0,1,...,N_{sym}-1\right\}}[n]$.
The baseband OFDM signal in the time domain can be written as~\cite{Heath:99}:
\vspace{-0.2cm}
\begin{equation}
\label{eqn:OFDM1}
s[n] = \frac {1} {\sqrt{N_{data}}} \displaystyle\sum\limits_{m=-\infty}^{\infty} {\displaystyle\sum\limits_{k=0}^{N_{data}-1}
									{a_{m,k}q[n-mN_{sym}]e^{j 2\pi k \frac {n-mN_{sym}} {N_{data}} }} }.
\vspace{-0.2cm}
\end{equation}
We assume that each data symbol $a_{m,k}$ is selected uniformly from a finite set of constellation points $\mA$, that satisfies a $180$ degrees symmetry. Thus, $\E\big\{a_{m,k}\big\}=0$. We also set $\E\big\{|a_{m,k}|^2\big\}=1$.
Letting each $a_{m,k}$ be selected in an i.i.d. manner over $k$ and $m$, it follows that $\E\big\{s[n]\} = 0$.
We denote the set of time indexes for which the corresponding signal samples are replicated into the cyclic prefix by $\mathcal{S}_{CP}$ and the set of time indexes for which the corresponding signal samples are cyclic prefix samples by $\mathcal{S}_{{CP}}'$. These are obtained as $\mathcal{S}_{CP} = \left\{\left.n\in \mathds{Z}\right| \left(n\mod N_{sym}\right) \geq N_{data} \right\}$ and $\mathcal{S}_{CP}' = \left\{\left.n\in \mathds{Z}\right| \left(n\mod N_{sym}\right) < N_{CP} \right\}$ respectively.
The autocorrelation function of the OFDM signal is~\cite[Eqn. (6)]{Heath:99}:
${c_{ss}}\left( {n,l} \right) = \delta \left[ l \right] + \delta \left[ {l - {N_{data}}} \right]{\bf{1}}_{\mathcal{S}_{{CP}}'}[n] + \delta \left[ {l + {N_{data}}} \right]{\bf{1}}_{\mathcal{S}_{{CP}}}[n]$.
Observe that both the mean function and autocorrelation function of the OFDM signal are periodic with respect to index $n$ with a period of $N_{sym}$. The OFDM signal is therefore wide-sense cyclostationary.
It should also be noted that for symmetric quadrature modulations, $\E\big\{a_{m,k}^2\} = 0$, therefore the conjugate autocorrelation of the OFDM signal is $c_{ss^*}(n,l) = \E\big\{s[n + l]s[n]\} = 0$.

The passband OFDM signal in the time domain can be generally written as~\cite[Ch. 2]{Nee:00} $d[n] = {\mathop{\rm Re}\nolimits} \left\{ {s[n]{e^{j2\pi {f_c}n{T_{samp}}}}} \right\}$, where $f_c$ denotes the carrier frequency. Note that as $\E\left\{s[n]\right\}=0$ then $\E\left\{d[n]\right\}=0$. For symmetric quadrature modulations, the autocorrelation function of the passband OFDM signal is:
\vspace{-0.2cm}
\begin{equation}
\label{eqn:OFDM_autocorr4}
{c_{dd}}\left( {n,l} \right) = \frac{1}{2}{\mathop{\rm Re}\nolimits} \left\{ {{c_{ss}}\left( {n,l} \right){e^{j2\pi {f_c}l{T_{samp}}}}} \right\},
\vspace{-0.2cm}
\end{equation}
where $T_{samp}$ denotes the sampling interval of the system. 
Observe that the passband OFDM signal is also cyclostationary with a period of $N_{sym}$.

\vspace{-0.4cm}
\subsection{A Cyclostationary Model For Narrowband PLC Noise}
\label{subsec:cyclo_noise}
\vspace{-0.25cm}
The PLC noise model proposed in~\cite{Katayama:06} incorporates all dominant narrowband noise components into a single mathematical model, which represents the noise as a real, passband, colored cyclostationary Gaussian process, $w[n]$, with a zero mean. 
At sample $n$ and frequency $f$, the power spectral density (PSD) $S(n,f)$ can be written as~\cite[Eqn. (11)]{Katayama:06} $S(n,f) = \beta[n]\alpha (f)$, where $\alpha (f)$ models the frequency dependence of the PSD, and is given by~\cite[Eqn. (12)]{Katayama:06} $\alpha (f) = \frac {\alpha_1} {2} e^{-\alpha_1 \left|f\right|}$, where the parameter $\alpha_1$ is chosen to fit the spectral properties of the measured colored noise.
In order to characterize $\beta[n]$, let the number of temporal noise components be $ L_{noise}$; the temporal behavior of the noise variance is expressed as a summation of the different noise components, that is~\cite[Eqn. (9)]{Katayama:06}:
$\beta[n] = \displaystyle\sum\limits_{i=0}^{L_{noise}-1}
				{A_i \left|\sin\left(\pi\frac{n}{N_{noise}}+\Theta _i\right)\right|^{n_i} }$, 					
where the parameters ${A_i}$, $n_i$ and $\Theta _i $ for $i = 0,1,\ldots, L_{noise}-1$, denote the characteristics of $i$-th noise component and $N_{noise}$ is the cyclic period derived from $N_{noise}=\frac{T_{AC}}{2T_{samp}}$, where $T_{AC}$ is the cycle duration of the mains voltage.
The autocorrelation function of the noise at sample index $n$, ${c_{ww}}(n,l)$, is obtained via the inverse Fourier transform of $S(n,f)$, see~\cite[Eqn. (17)]{Katayama:06}.

Another relevant cyclostationary noise model is proposed in~\cite{Nassar:12}.
In this model the cyclostationary noise is obtained as the output of an LPTV system when the input is a white Gaussian stochastic process (WGSP). The work in \cite{Nassar:12} models the cyclostationary noise by dividing $N_{noise}$ into $M$ time intervals and filtering a WGSP with a finite set of $M$ LTI filters in parallel. At each time interval, the noise signal is taken from the output of one of the filters, and the selected filter changes periodically. The noise samples are therefore obtained as
$w[n] = \sum\limits_{l =  - \infty }^\infty  {h[n,l]\upsilon [l]}  = \sum\limits_{i = 1}^M {{{\bf{1}}_{n \in {R_i}}}} \sum\limits_{l =  - \infty }^\infty  {{h_i}[n - l]\upsilon [l]}$,
where $h[n,l]$ denotes the LPTV filter realized using a set of $M$ LTI filters $h_i[l]$, $i=1,2,\ldots,M$, and $R_i$, $i=1,2,\ldots,M$, denotes the time indexes in which the noise samples are taken as the output of the LTI filter $h_i[l]$. Lastly, $\upsilon [n]$ is a zero mean, unit variance WGSP. $w[n]$ is clearly cyclostationary since the PDF of $w[n]$, denoted $p_{w[n]}(z)$, satisfies $p_{w[n]}(z) = p_{w[n+kN_{noise}]}(z)$ for every integer $k$.

In the simulations section we demonstrate the performance improvement for both noise models~\cite{Katayama:06} and~\cite{Nassar:12}.

\vspace{-0.4cm}
\subsection{Frequency Shift Filtering}
\label{subsec:FRESH}
\vspace{-0.25cm}
The cyclostationary equivalent of  linear time-invariant filtering is the frequency shift (FRESH) filtering. In~\cite{Gardner:93}, linear-conjugate-linear (LCL) FRESH filtering is developed. The LCL FRESH filter is a time varying linear filter represented by the impulse responses $h[n,l]$ and $h^c[n,l]$. The output signal $y[n]$ for input $r[n]$ is given by $y[n] = \sum\limits_{l=-\infty}^{\infty} {h[n,l]r[l] + } \sum\limits_{m=-\infty}^{\infty} {{h^c}[n,m]{r^*}[m]}$.

The impulse response $h[n,l]$ is defined by $h[n,l] = \sum\limits_{k=0}^{N_0-1} {\widetilde{h}_k} [n - l]{e^{-j2\pi{\alpha _k}l}}$, 
where $N_0$ is the cyclic period of the FRESH filter and ${\alpha _k} = \frac{k}{{{N_{0}}}}$ is the corresponding cyclic frequency.
A similar relationship holds for ${h}^c[n,l]$ and $\widetilde{h}_k^c[l]$.
The relationship between the input $r[n]$  and the output $y[n]$ of the filter can be therefore written as
\vspace{-0.2cm}
\begin{equation}
\label{eqn:FRESH_Input_Output}
y[n] = \sum\limits_{k = 0}^{{N_0} - 1} {\left( {\sum\limits_{l =  - \infty }^\infty  {{{\widetilde h}_k}\left[ {n - l} \right]{r_k}\left[ l \right]}  + \sum\limits_{l =  - \infty }^\infty  {\widetilde h_k^c\left[ {n - l} \right]r_k^c\left[ l \right]} } \right)} ,
\vspace{-0.2cm}
\end{equation}
where ${r_k}[n] = r[n]{e^{-j2\pi \alpha_k n}}$ and $r_k^c[n] = {r^*}[n]{e^{-j2\pi \alpha_k n}}$. 
From \eqref{eqn:FRESH_Input_Output} we observe that the system performs linear time invariant (LTI) filtering of frequency shifted versions of $r[n]$, therefore, the FRESH filter can be modeled as an LTI filter-bank  applied to the frequency shifted versions of the input signal~\cite{Ojeda:11},~\cite{Zhang:95}. 
The optimal FRESH filter in the sense of minimal TA-MSE between the output signal and the desired signal is developed in~\cite{Gardner:93}.
For cyclostationary signals with zero conjugate cyclic autocorrelation, the LCL FRESH filter specializes the linear FRESH filter.

Consider the received signal $r[n]=d[n]+w[n]$, where $d[n]$ denotes the desired cyclostationary signal and $w[n]$ denotes the additive noise. Let each LTI filter in the implementation of the FRESH filter consist of a finite impulse response (FIR) filter of $L_{\mathrm{FIR}}$ taps.
Let $K$ denote the number of cyclic frequencies used by the FRESH filter, ${\alpha _k}$ denote the $k$-th cyclic frequency, $h_k[i] = \widetilde{h}_k^*[i]$ denote the conjugate of the $i$-th coefficient of the $k$-th FIR,  ${\bf{z}}[n]$ denote the frequency shifted input vector at time $n$, defined as:
${\bf{z}}[n] = \left[{\bf{r}}_0[n],{\bf{r}}_1[n],...,{\bf{r}}_{K-1}[n]\right]^T$,
with $\left({\bf{r}}_k[n]\right)_i = r[n-i]e^{-j2\pi \alpha _k \left(n - i\right)}$, $i
\in \mathcal{L} \triangleq \big\{0,1,\ldots,L_{\mathrm{FIR}}-1\big\}$.
Finally, let $\bf{h}$ denote the concatenated conjugate of the FIR coefficients vectors obtained by ${\bf{h}} = [{\bf{h}}_0,{\bf{h}}_1,...,{\bf{h}}_{K-1}]^T$, where $\left({\bf{h}}_k\right)_i = h_k[i]$, $i \in \mathcal{L}$.
The input-output relationship of the FRESH filter can now be written as
\vspace{-0.2cm}
\begin{equation}
\label{eqn:FRESH_Linear_Input_Output}
y[n]	= \sum\limits_{k = 0}^{K-1} {\sum\limits_{i = 0}^{L_{\mathrm{FIR}} - 1} {h_k^*[i]r[n - i]{e^{-j2\pi {\alpha _k}(n - i)}}} }
 			= {{\bf{h}}^H}{\bf{z}}\left[n\right].
\vspace{-0.2cm}
\end{equation}
As detailed in~\cite{Gardner:93},~\cite{Ojeda:11},
the optimal FRESH filter is obtained as:
\vspace{-0.2cm}
\begin{equation}
\label{eqn:FRESH_Linear_Optimal_H1}
{\bf{h}} = {\bar{\bf{C}}_{{\bf{zz}}}}^{ - 1}{\bar{\bf{c}}_{{\bf{z}}d}},
\vspace{-0.2cm}
\end{equation}
where ${\bar{\bf{c}}_{{\bf{z}}d}}\triangleq\left\langle {\bf{c}}_{{\bf{z}}d}[n]\right\rangle=\left\langle \E\big\{{\bf{z}}[n]d^*[n]\}\right\rangle=\frac{1}{N_0}\sum\limits_{n=0}^{N_0-1}{\E\big\{{\bf{z}}[n]d^*[n]\}}$ denotes the time-averaged cross-correlation vector of the desired signal and the frequency shifted received vector, ${\bar{\bf{C}}_{\bf{z}\bf{z}}}\triangleq\left\langle {\bf{C}}_{\bf{z}\bf{z}}[n]\right\rangle=\left\langle \E\big\{{\bf{z}}[n]{\bf{z}}^H[n]\}\right\rangle$, and $L_{\mathrm{FIR}}$ is at least as large as the largest value of $l\in\big\{0,1,\ldots,N_0-1\big\}$ for which exists an index $n$ such that $c_{dd}(n,l) \neq 0$. Note that all time averaging is over some $N_0$ which is an integer multiple of the period of the desired signal $d[n]$.

Next, following~\cite{Chen:11}, we use the independence of the desired signal and the noise, together with the fact that $\E\left\{w[n]\right\}=0$, to write ${\bf{c}}_{{\bf{z}}d}[n] =\big[c_{dd}(n,0){e^{ - j2\pi {\alpha _0}n}},$ $ c_{dd}(n-1,0){e^{ - j2\pi {\alpha _0}\left(n-1\right)}},$ $\ldots,$ $c_{dd}(n,- L_{\mathrm{FIR}} + 1)e^{ - j2\pi {\alpha _0}(n - L_{\mathrm{FIR}} + 1)},$
$c_{dd}(n,0)e^{ - j2\pi {\alpha _1}n},$ $\ldots,$ $c_{dd}(n,- L_{\mathrm{FIR}} + 1)e^{ - j2\pi {\alpha _{K-1}}(n - L_{\mathrm{FIR}} + 1)}\big]^T$.
Consider ${\left( {{\bf{c}}_{{\bf{z}}d}}[n] \right)_i}$, the $i$-th element of the vector ${\bf{c}}_{{\bf{z}}d}^{}[n] $; by writing the index $i$ as $i = p_iL_{\mathrm{FIR}}+q_i$, ${q_i} \in \mathcal{L}$, ${p_i} \in \mathcal{K} \triangleq \big\{ 0,1,...,K - 1 \big\}$, we can write
\vspace{-0.2cm}
\begin{equation}
\label{eqn:FRESH_Linear_Optimal_Rzd}
{\left( {{\bf{c}}_{{\bf{z}}d}}[n] \right)_i} = {c_{dd}}(n, -{q_i}){e^{ - j2\pi {\alpha _{{p_i}}}(n - {q_i})}},
\vspace{-0.2cm}
\end{equation}
$i \in \mathcal{M} \triangleq \big\{0,1,\dots,KL_{\mathrm{FIR}}-1\big\}$,
and the $i$-th component of the time-averaged vector ${{\bar{\bf{c}}_{{\bf{z}}d}}}$ can be expressed as ${\left({{\bar{\bf{c}}_{{\bf{z}}d}}}\right)_i}= \left\langle {\left( {{\bf{c}}_{{\bf{z}}d}}[n] \right)_i}\right\rangle$.
Next, consider the autocorrelation matrix ${{\bf{C}}_{{\bf{zz}}}}[n]$: Writing the indexes $u,v\in \mathcal{M}$ as $u = p_uL_{\mathrm{FIR}}+q_u$ and $v = p_vL_{\mathrm{FIR}}+q_v$, ${p_u},{p_v} \in \mathcal{K}$, ${q_u},{q_v} \in \mathcal{L}$, the element at the $u$-th row and $v$-th column of ${\bf{C}}_{{\bf{zz}}}[n]$ may be expressed as 
\vspace{-0.2cm}
\begin{align}	
\label{eqn:FRESH_Linear_Optimal_Rzz}
{\left( {{\bf{C}}_{{\bf{zz}}}}[n] \right)_{u,v}} &= \E\big\{ r[n - {q_u}]{e^{ - j2\pi {\alpha _{{p_u}}}(n - {q_u})}}{r^*}[n - {q_v}]{e^{j2\pi {\alpha _{{p_v}}}(n - {q_v})}}\} \notag \\
	&= {c_{dd}}(n - {q_v},{q_v} - {q_u}){e^{ - j2\pi {\alpha _{{p_u}}}(n - {q_u})}}{e^{j2\pi {\alpha _{{p_v}}}(n - {q_v})}} \notag \\		&\qquad + {c_{ww}}(n - {q_v},{q_v} - {q_u}){e^{ - j2\pi {\alpha _{{p_u}}}(n - {q_u})}}{e^{j2\pi {\alpha _{{p_v}}}(n - {q_v})}},
\vspace{-0.2cm}	
\end{align}
and ${\left( {{\bar{\bf{C}}_{{\bf{zz}}}}} \right)_{u,v}}= \left\langle {\left( {{\bf{C}}_{{\bf{zz}}}}[n] \right)_{u,v}}\right\rangle$.
Since the output signal produced by the cyclic Wiener filter is orthogonal to the error~\cite{Gardner:86}, the TA-MSE between the output and the desired signal can be written as~\cite[Pg. 431]{Gardner:86}
$ \mbox{TA-MSE} = \left\langle \E\left\{ {\left|y[n] - d[n]\right|^2}\right\}\right\rangle  = {P_d} - {\bar{\bf{c}}_{{\bf{z}}d}}^H{\bar{\bf{C}}_{{\bf{zz}}}}^{ - 1}{\bar{\bf{c}}_{{\bf{z}}d}}$,
where ${P_d}$ denotes the average energy of the desired signal evaluated as
${P_d} = \left\langle \E\big\{ d[n]{d^*}[n]\}\right\rangle  = \left\langle {c_{dd}}(n,0)\right\rangle$.

\vspace{-0.25cm}
\section{Minimum TA-MSE Signal Recovery}
\label{sec:Model}
\vspace{-0.25cm}
In this section we present a new receiver scheme for the recovery of an OFDM signal received over an additive cyclostationary noise channel. The scheme exploits the spectral correlation of the OFDM signal $d[n]$ {\em as well as the spectral correlation of the noise}.
The received signal is given by $r[n]=d[n]+w[n]$, where $w[n]$ is the noise, and $d[n]$ and $w[n]$ are mutually independent. $N_{sym}$ and $N_{noise}$ denote the periods of $d[n]$ and $w[n]$, respectively.
All time averages denoted $\left\langle \cdot\right\rangle$ are over the least common multiple of $N_{sym}$ and $N_{noise}$.

\vspace{-0.4cm}
\subsection{A New Receiver Algorithm: Signal Recovery with Noise Estimation and Cancellation}
\label{subsec:Noise_Estimate}
\begin{figure}
	\centering
		\includegraphics[width=0.75\textwidth]{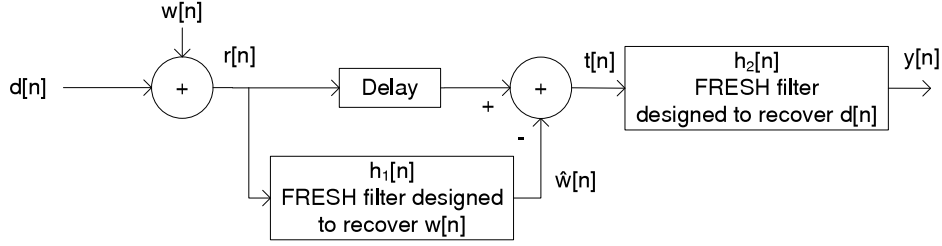}
\vspace{-0.4cm}
	\caption{FRESH filtering for signal recovery with noise extraction and cancellation.}
	\label{fig:SystemModel2}
	\vspace{-0.8cm}
\end{figure}

\vspace{-0.25cm}
Our new receiver algorithm exploits the cyclostationary properties of both the OFDM signal and the noise by applying noise estimation and cancellation prior to signal extraction. The main novelty of the scheme is the utilization of the cyclostationary properties of the noise as well as those of the signal, in contrast to using only the properties of the information signal, which is the approach of previous schemes. The algorithm processing, depicted in Fig. \ref{fig:SystemModel2}, consists of two FRESH filters in series:
the first filter, $h_1[n]$, is a noise estimation FRESH filter tuned to extracting the cyclostationary noise. The estimated noise is then subtracted from the received signal, and then a signal extraction FRESH filter, $h_2[n]$, tuned to recovering the OFDM signal is applied.
We expect that this structure shall have superior performance at low SNR, which is the relevant operating regime for narrowband PLC~\cite{Pavlidou:03}.
Since we operate in passband, both the signal and the noise are real-valued, thus the LCL FRESH filters, $h_1[n]$ and $h_2[n]$, consist only of linear FRESH filters.

Let $K_1$ be the number of cyclic frequencies used by $h_1[n]$,
$L_{\mathrm{FIR1}}$ be the length of the FIR filter at each branch of $h_1[n]$, and $\alpha _k=\frac{k}{N_{noise}}$ denote the $k$-th cyclic frequency.
From \eqref{eqn:FRESH_Linear_Optimal_H1}, the FRESH filter $h_1[n]$, designed to recover the noise, is obtained as
\vspace{-0.2cm}
\begin{equation}
\label{eqn:Model2_FRESH_H1_1}
{\bf{h}}_1^{} = {\bar{\bf{C}}_{{\bf{rr}}}}^{ - 1}{{\bar{\bf{c}}_{{\bf{r}}w}}},
\vspace{-0.2cm}
\end{equation}
where ${\bf{r}}[n] = \left[{\bf{r}}_0[n],{\bf{r}}_1[n],...,{\bf{r}}_{K_1-1}[n]\right]^T$, $\left({\bf{r}}_k[n]\right)_i = r[n-i]e^{-j2\pi \alpha _k \left(n-i\right)}$, $k \in \mathcal{K}_1 \triangleq \big\{0,1,\ldots,K_1-1\big\}$, $i \in \mathcal{L}_1 \triangleq \big\{0,1,\ldots,L_{\mathrm{FIR1}}-1\big\}$.
${\bar{\bf{C}}_{{\bf{rr}}}} = \left\langle {\bf{C}}_{{\bf{rr}}}[n]\right\rangle$, with ${\bf{C}}_{{\bf{rr}}}[n] = \E\big\{{\bf{r}}[n]{\bf{r}}^H[n] \big\}$, and ${\bar{\bf{c}}_{{\bf{r}}w}}=\left\langle {\bf{c}}_{{\bf{r}}w}[n]\right\rangle$, with
${\bf{c}}_{{\bf{r}}w}^{}[n] = \E\left\{ {\bf{r}}[n] {w^*}[n]\right\}$.
We let the indexes $u,v\in \mathcal{M}_1 \triangleq \big\{0,1,\ldots,K_1L_{\mathrm{FIR1}}-1 \big\}$ be written as $u = p_uL_{\mathrm{FIR1}}+q_u$ and $v = p_vL_{\mathrm{FIR1}}+q_v$, ${p_u},{p_v} \in \mathcal{K}_1$, ${q_u},{q_v} \in \mathcal{L}_1$. Applying the steps used in the derivation of \eqref{eqn:FRESH_Linear_Optimal_Rzd} and \eqref{eqn:FRESH_Linear_Optimal_Rzz}, we have
\vspace{-0.2cm}
\begin{align}
\label{eqn:FRESH_Linear_Optimal_Rrr}
{\left( {{\bf{C}}_{{\bf{rr}}}}[n] \right)_{u,v}} &=
	{c_{dd}}(n - {q_v},{q_v} - {q_u}){e^{ - j2\pi {\alpha _{{p_u}}}(n - {q_u})}}{e^{j2\pi {\alpha _{{p_v}}}(n - {q_v})}} \notag \\		&\qquad + {c_{ww}}(n - {q_v},{q_v} - {q_u}){e^{ - j2\pi {\alpha _{{p_u}}}(n - {q_u})}}{e^{j2\pi {\alpha _{{p_v}}}(n - {q_v})}},
\vspace{-0.2cm}
\end{align}
and
\vspace{-0.2cm}
\begin{equation}
\label{eqn:Model2_FRESH_H1_Rzw2}
{\left( {{\bf{c}}_{{\bf{r}}w}}[n] \right)_u}=c_{ww}\left({n,-q_u}\right){e^{-j2\pi {\alpha _{p_u}}(n - q_u)}},
\vspace{-0.2cm}
\end{equation}
where ${c_{dd}}(n,l)$ is obtained from \eqref{eqn:OFDM_autocorr4} and ${c_{ww}}(n,l)$ is specified by the noise model, e.g.,~\cite{Katayama:06} or~\cite{Nassar:12}. The estimated noise is therefore $\hat w[n] = {\bf{h}}_1^H{\bf{r}}[n]$.

Consider next the FRESH filter $h_2[n]$, designed to recover the OFDM signal $d[n]$. The input signal to $h_2[n]$ consists of the received signal after the estimated noise was subtracted, that is $t[n] = r[n] - \hat w[n] = d[n] + w[n] - \hat w[n]$. Note that in practice all filters are causal and therefore appropriate delays must be introduced in $w[n]$ and in $d[n]$, in the expression for $t[n]$ (see Fig \ref{fig:SystemModel2}). To avoid cluttering the notation we derive non-causal versions of the filter, but as all filters are FIR, introducing  the appropriate delays is simple in a practical setup.
From \eqref{eqn:FRESH_Linear_Optimal_H1}, the FRESH filter $h_2[n]$ is obtained as
\vspace{-0.2cm}
\begin{equation}
\label{eqn:Model2_FRESH_H2_1}
{\bf{h}}_2^{} = {\bar{\bf{C}}_{{\bf{tt}}}}^{ - 1}{{\bar{\bf{c}}_{{\bf{t}}d}}},
\vspace{-0.2cm}
\end{equation}
where ${\bar{\bf{c}}_{{\bf{t}}d}}=\left\langle {\bf{c}}_{{\bf{t}}d}[n]\right\rangle=\left\langle \E\big\{{\bf{t}}[n]d^*[n]\}\right\rangle$, ${\bar{\bf{C}}_{\bf{t}\bf{t}}}=\left\langle {\bf{C}}_{\bf{t}\bf{t}}[n]\right\rangle=\left\langle \E\big\{{\bf{t}}[n]{\bf{t}}^H[n]\}\right\rangle$, the vector ${\bf{t}}[n]$ is obtained by ${\bf{t}}[n] = \left[{\bf{t}}_0[n], {\bf{t}}_1[n], \ldots, {\bf{t}}_{K_2-1}[n]\right]^T$, where $K_2$ denotes the number of cyclic frequencies used by $h_2[n]$,
$\left({\bf{t}}_k[n]\right)_i = t[n-i]e^{-j2\pi \beta _k \left(n-i\right)}$, $i \in \mathcal{L}_2 \triangleq \big\{0,1,\ldots,L_{\mathrm{FIR2}}-1\big\}$,
$L_{\mathrm{FIR2}}$ is the length of the FIR filter at each branch of  $h_2[n]$, and $\beta _k=\frac{k}{N_{sym}}$ denotes the $k$-th cyclic frequency used in $h_2[n]$.
In order to evaluate ${{\bf{C}}_{{\bf{tt}}}}[n]$, we let the indexes $u,v\in \mathcal{M}_2 \triangleq \big\{0,1,\ldots,K_2L_{\mathrm{FIR2}}-1 \big\}$ be written as $u = p_uL_{\mathrm{FIR2}}+q_u$ and $v = p_vL_{\mathrm{FIR2}}+q_v$, ${q_u},{q_v} \in \mathcal{L}_2$, ${p_u},{p_v} \in \mathcal{K}_2 \triangleq \big\{ {0,1,\ldots,K_2 - 1} \big\}$. Then, we have ${\big( {{{\bf{C}}_{{\bf{tt}}}}}[n] \big)_{u,v}} = \E\big\{ {t[n - {q_u}]{e^{ - j2\pi {\beta _{{p_u}}}(n - {q_u})}}{t^*}[n - {q_v}]{e^{j2\pi {\beta _{{p_v}}}(n - {q_v})}}} \big\} $, thus
\vspace{-0.2cm}
\begin{align}
\label{eqn:Model2_FRESH_H2_Rtt1}
{\left( {{{\bf{C}}_{{\bf{tt}}}}}[n] \right)_{u,v}}
	&= \E\big\{ {t[n - {q_u}]{e^{ - j2\pi {\beta _{{p_u}}}(n - {q_u})}}{t^*}[n - {q_v}]{e^{j2\pi {\beta _{{p_v}}}(n - {q_v})}}} \big\} \notag \\
	&= {c_{tt}}\left( {n - {q_v},{q_v} - {q_u}} \right){e^{j2\pi \left( {{\beta _{{p_v}}}(n - {q_v}) - {\beta _{{p_u}}}(n - {q_u})} \right)}}.
\vspace{-0.2cm}
\end{align}

Next, we define $ {\bf{d}}[n]  \buildrel \Delta \over = \left[{\bf{d}}_0[n],{\bf{d}}_1[n],\ldots,{\bf{d}}_{K_1-1}[n]\right]^T$,
where $\left({\bf{d}}_k[n]\right)_i = d[n-i]e^{-j2\pi \alpha _k \left(n-i\right)}$, $i \in \mathcal{L}_1$, and ${\bf{w}}[n] \buildrel \Delta \over = \big[{\bf{w}}_0[n],{\bf{w}}_1[n],$$\ldots,$${\bf{w}}_{K_1-1}[n]\big]^T$, where
$\left({\bf{w}}_k[n]\right)_i = w[n-i]e^{-j2\pi \alpha _k \left(n-i\right)}$, $i \in \mathcal{L}_1$.
${\bf{r}}[n]$ can now be written as ${\bf{r}}[n] = {\bf{d}}[n] + {\bf{w}}[n]$.
Now, let us denote by ${d_1}[n] = {\bf{h}}_1^H{\bf{d}}[n]$ the desired signal component at the output of $h_1[n]$, and by ${w_1}[n] = {\bf{h}}_1^H{\bf{w}}[n]$ the noise component at the output  of $h_1[n]$. The input signal of $h_2[n]$ may therefore be expressed as $t[n] =  d[n] + w[n] - {d_1}[n] - {w_1}[n] = {d_2}[n] + {w_2}[n]$, where ${d_2}[n] = d[n] - {d_1}[n]$ and $ {w_2}[n] = w[n] - {w_1}[n]$. Note that, since $d[n]$ and $w[n]$ are mutually independent, then $d_2[n]$ and $w_2[n]$ are mutually independent.
$c_{tt}(n,l)$ may therefore be obtained by:
\vspace{-0.2cm}
\begin{equation}
\label{eqn:Model2_FRESH_H2_Rtt2}
{c_{tt}}\left( {n,l} \right) = \E\left\{ {t[n + l]{t^*}[n]} \right\}
= {c_{d_2d_2}}\left( {n,l} \right) + {c_{w_2w_2}}\left( {n,l} \right).
\vspace{-0.2cm}
\end{equation}

As the noise models of both~\cite{Katayama:06} and~\cite{Nassar:12} include a stationary component, we have $
c_{ww}^0(l) = \frac{1}{{{N_{noise}}}}\sum\limits_{n = 0}^{{N_{noise}} - 1} {{c_{ww}}(n,l) \ne 0}$, it follows from~\cite{Gardner:93}, that the FRESH filter designed to recover the noise must include the cyclic frequency ${\alpha _{{k_0}}} = 0$, $k_0 \in \mathcal{K}_1$.
Let ${{\bf{i}}_{{L_{\mathrm{FIR1}}} \cdot {k_0}}}$  denote a column vector such that its $i$-th coordinate is obtained by $
{\left( {{\bf{i}}_{{L_{\mathrm{FIR1}}} \cdot {k_0}}} \right)_i} = \delta [i - {L_{\mathrm{FIR1}}} \cdot {k_0}]$.
We may now write the data signal sample as $d[n] = {\bf{i}}_{{L_{\mathrm{FIR1}}} \cdot {k_0}}^H{\bf{d}}[n]$, and write the noise sample as $w[n] = {\bf{i}}_{{L_{\mathrm{FIR1}}} \cdot {k_0}}^H{\bf{w}}[n]$.
Next, we write ${d_2}[n] = d[n] - {{\bf{h}}_1^H{\bf{d}}} [n] = {{\check{\bf i}_1^H}{\bf{d}}}[n]$, where $\check{\bf i}_1 \triangleq {\bf{i}}_{{L_{\mathrm{FIR1}}} \cdot {k_0}}^{} - {\bf{h}}_1$.
Letting ${{\bf{C}}_{{\bf{dd}}}}(n,l) = {\E\left\{ {{\bf{d}}[n + l]{{\bf{d}}^H}[n]} \right\}}$, the autocorrelation of ${d_2}[n]$ may therefore be expressed as
\vspace{-0.2cm}
\begin{equation}
\label{eqn:Model2_SNR_out_Rd2d2_6}
{c_{{d_2}{d_2}}}\left( {n,l} \right) = \E\left\{ {{d_2}[n + l]d_2^*[n]} \right\}
  = {\check{\bf i}_1^H}{{\bf{C}}_{{\bf{dd}}}}(n,l)\check{\bf i}_1.
  \vspace{-0.2cm}
\end{equation}
By writing the indexes $u_1,v_1\in \mathcal{M}_1$ as $u_1 = p_{u_1}L_{\mathrm{FIR1}}+q_{u_1}$ and $v_1 = p_{v_1}L_{\mathrm{FIR1}}+q_{v_1}$, where ${p_{u_1}},{p_{v_1}} \in  \mathcal{K}_1$, and ${q_{u_1}},{q_{v_1}} \in  \mathcal{L}_1$, we write ${\left( {{\bf{C}}_{{\bf{dd}}}}(n,l) \right)_{{u_1},{v_1}}} = \E\big\{ d[n + l - {q_{{u_1}}}]{e^{ - j2\pi {\alpha _{{p_{{u_1}}}}}(n + l - {q_{{u_1}}})}}d_{}^*[n - {q_{{v_1}}}]{e^{j2\pi {\alpha _{{p_v}_{_1}}}(n - {q_{{v_1}}})}} \big\} $, therefore
\vspace{-0.2cm}
\begin{align}
\label{eqn:Model2_SNR_out_Rd2d2_7}
{\left( {{\bf{C}}_{{\bf{dd}}}}(n,l) \right)_{{u_1},{v_1}}} 
	&= \E\left\{ {d[n + l - {q_{{u_1}}}]{e^{ - j2\pi {\alpha _{{p_{{u_1}}}}}(n + l - {q_{{u_1}}})}}d_{}^*[n - {q_{{v_1}}}]{e^{j2\pi {\alpha _{{p_v}_{_1}}}(n - {q_{{v_1}}})}}} \right\} \notag \\
	&= {c_{dd}}\left( {n - {q_{{v_1}}},{q_{{v_1}}} + l - {q_{{u_1}}}} \right){e^{j2\pi \left( {{\alpha _{{p_{{v_1}}}}}(n - {q_{{v_1}}}) - {\alpha _{{p_{{u_1}}}}}(n + l - {q_{{u_1}}})} \right)}}.
\vspace{-0.2cm}
\end{align}
Applying the steps used in the derivation of ${{{{c}}_{{{{d}}_2}{{{d}}_2}}}}(n,l)$ to the derivation of ${{{{c}}_{{{{w}}_2}{{{w}}_2}}}}(n,l)$, we obtain
\vspace{-0.2cm}
\begin{equation}
\label{eqn:Model2_SNR_out_Rw2w2_2}
{c_{{w_2}{w_2}}}\left( {n,l} \right) = {\check{\bf i}_1^H}{{\bf{C}}_{{\bf{ww}}}}(n,l) \check{\bf i}_1,
\vspace{-0.2cm}
\end{equation}
where
\vspace{-0.2cm}
\begin{equation}
\label{eqn:Model2_SNR_out_Rw2w2_4}
{\left( {{\bf{C}}_{{\bf{ww}}}}(n,l) \right)_{{u_1},{v_1}}} 
= {c_{ww}}\left( {n - {q_{{v_1}}},{q_{{v_1}}} + l - {q_{{u_1}}}} \right){e^{j2\pi \left( {{\alpha _{{p_{{v_1}}}}}(n - {q_{{v_1}}}) - {\alpha _{{p_{{u_1}}}}}(n + l - {q_{{u_1}}})} \right)}}.
\vspace{-0.2cm}
\end{equation}
The correlation \eqref{eqn:Model2_FRESH_H2_Rtt1} is obtained by plugging \eqref{eqn:Model2_SNR_out_Rd2d2_6}, \eqref{eqn:Model2_SNR_out_Rd2d2_7}, \eqref{eqn:Model2_SNR_out_Rw2w2_2}, and  \eqref{eqn:Model2_SNR_out_Rw2w2_4} into \eqref{eqn:Model2_FRESH_H2_Rtt2}, and plugging \eqref{eqn:Model2_FRESH_H2_Rtt2} into \eqref{eqn:Model2_FRESH_H2_Rtt1}.

Next, ${{\bf{c}}_{{\bf{t}}d}}[n]$ may be expressed as ${\bf{c}}_{{\bf{t}}d}^{}[n] = \E\left\{ {\bf{t}}[n]{d^*}[n]\right\}$.
We note that $\E\{ t[n + l]{d^*}[n]\}  = \E\{ \left( {d_2[n + l] + w_2[n + l] } \right){d^*}[n]\}  = {c_{d_2d}}\left( {n,l} \right)$.
Therefore, by writing the index $i$ as $i = p_iL_{\mathrm{FIR2}}+q_i$, ${p_i} \in \mathcal{K}_2$, ${q_i} \in \mathcal{L}_2$, we obtain
\vspace{-0.2cm}
\begin{equation}
\label{eqn:Model2_FRESH_H2_Rtd3}
{\left( {{\bf{c}}_{{\bf{t}}d}}[n] \right)_i}=c_{d_2d}\left({n,-q_i}\right){e^{-j2\pi {\beta _{p_i}}(n - q_i)}}, \qquad i \in \mathcal{M}_2 .
\vspace{-0.2cm}
\end{equation}
Note that ${c_{{d_2}d}}(n,l) = \E\left\{ {{d_2}[n + l]{d^*}[n]} \right\}
			= {\check{\bf i}_1^H}\E\left\{ {{\bf{d}}[n + l]{d^*}[n]} \right\}$.
By writing the index $u\in \mathcal{M}_1$ as $u = p_{u}L_{\mathrm{FIR1}}+q_{u}$, ${p_u} \in \mathcal{K}_1$, ${q_u} \in \mathcal{L}_1$,  we have
${\left( \E\left\{ {{\bf{d}}[n + l]{d^*}[n]} \right\} \right)_u} = \E\big\{ d[n + l - {q_u}]e^{ - j2\pi {\alpha _{{p_u}}}\left( {n + l - {q_u}} \right)}{d^*}[n] \big\} = {c_{dd}}(n,l - {q_u})e^{ - j2\pi {\alpha _{{p_u}}}\left( {n + l - {q_u}} \right)}$.
Equations \eqref{eqn:FRESH_Linear_Optimal_Rrr}, \eqref{eqn:Model2_FRESH_H1_1}, and \eqref{eqn:Model2_FRESH_H1_Rzw2} provide a closed form expression for $h_1[n]$, and equations \eqref{eqn:Model2_FRESH_H2_1}, \eqref{eqn:Model2_FRESH_H2_Rtt1}, and \eqref{eqn:Model2_FRESH_H2_Rtd3} provide a closed form expression for $h_2[n]$.
The TA-MSE of the proposed receiver is given by:
\vspace{-0.2cm}
\begin{equation}
\label{eqn:model2_output2}	
\mbox{TA-MSE} = {{{P_d} - {\bar{\bf{c}}_{{\bf{t}}d}}^H{\bar{\bf{C}}_{{\bf{tt}}}}^{ - 1}{{\bar{\bf{c}}_{{\bf{t}}d}}}}}.
\vspace{-0.2cm}
\end{equation}

\vspace{-0.4cm}
\subsection{Best of Previous Work: A FRESH Filter Designed in~\cite{Chen:11} for Direct Signal Recovery}
\label{subsec:Signal_Extract}
\vspace{-0.25cm}
The best previously proposed scheme for this model is a FRESH filter tuned to extracting the OFDM signal based on the minimum TA-MSE criterion, proposed in~\cite{Chen:11}. Note that for passband OFDM signals and for baseband OFDM signals which employ a quadrature constellation (e.g., QPSK, QAM) for modulating the subcarriers, the conjugate cyclic autocorrelation is zero, as shown in Subsection \ref{subsec:cyclo_OFDM}, and as a result the LCL FRESH filter specializes to a linear FRESH filter.

\begin{figure}
	\centering
		\includegraphics[width=0.75\textwidth]{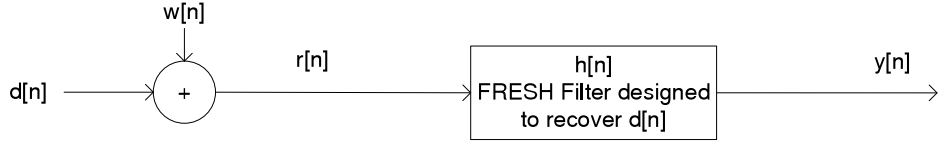}
\vspace{-0.5cm}
	\caption{A block diagram for the system of~\cite{Chen:11}: A FRESH filter performing direct signal recovery.}
	\label{fig:SystemModel1}
	\vspace{-0.8cm}
\end{figure}

The signal and system model used in~\cite{Chen:11} are depicted in Fig. \ref{fig:SystemModel1}.
The filter $h[n]$ is derived using \eqref{eqn:FRESH_Linear_Optimal_H1}, where ${\bf{c}}_{{\bf{z}}d}[n]$ is calculated via \eqref{eqn:FRESH_Linear_Optimal_Rzd}, and ${\bf{C}}_{{\bf{zz}}}[n]$ is calculated using \eqref{eqn:FRESH_Linear_Optimal_Rzz}.
The TA-MSE of the proposed model is given by ${{\mbox{TA-MSE}}} = {{{P_d} - {\bar{\bf{c}}_{{\bf{z}}d}}^H{\bar{\bf{C}}_{{\bf{zz}}}}^{ - 1}{{\bar{\bf{c}}_{{\bf{z}}d}}}}}$.

\vspace{-0.4cm}
\subsection{Evaluating the Impact of Output Scaling on the Performance}
\label{subsec:Scaling}
\vspace{-0.25cm}
As discussed in~\cite[Ch. 2.4]{Kay:93}, filters designed according to the minimum MSE criterion may induce a bias at the output of the filter. Since the coefficients of both FRESH filters in our proposed algorithm are selected to minimize the TA-MSE, the recovered OFDM signal at the output of the receiver suffers from a scaling effect which is analyzed in the following.

Consider the receiver depicted in Fig. \ref{fig:SystemModel2}. For a given index $n$, we define the scaling of the output of the filter relative to the desired signal $d[n]$ as
$\psi [n] = \E\left\{ \left. \frac{{ {y[n]} }}{{d[n]}} \right| d[n]\right\} = \frac{{\E\left\{ {y[n]|d[n]} \right\}}}{{d[n]}}$. Let $h_{v,k}[i]$ denote the $i$-th coefficient of the FIR filter of the $k$-th branch of $h_v[n]$, $v\in\big\{1,2\big\}$, $k\in\big\{0,1,\ldots,K_v - 1\big\}$ and $i\in\big\{0,1,\ldots,{L_{{\mathrm{FIR}}v}} - 1\big\}$.
Using \eqref{eqn:FRESH_Linear_Input_Output} we write
$\E\left\{ {y[n]|d[n]} \right\} = \E\Big\{ { \sum\limits_{k = 0}^{K_2 - 1} {\sum\limits_{i = 0}^{{L_{\mathrm{FIR2}}} - 1} {h_{2,k}^*[i]t[n - i]{e^{ - j2\pi {\beta _k}n}}} } \Big|d[n]} \Big\}$. Recalling that $t[n] = r[n] - \hat w[n]$, we write $y[n] = \sum\limits_{k = 0}^{{K_2} - 1} {}\sum\limits_{i = 0}^{{L_{\mathrm{FIR2}}} - 1} {}h_{2,k}^*[i]\left( {r[n - i] - \hat w[n - i]} \right){e^{ - j2\pi {\beta _k}n}}$. Since $\hat w[n] = {\bf{h}}_1^H{\bf{r}}[n]$, we obtain $y[n] = $ \\ $ \sum\limits_{k = 0}^{{K_2} - 1} {}\sum\limits_{i = 0}^{{L_{\mathrm{FIR2}}} - 1} {}h_{2,k}^*[i]\Big( r[n - i] - \sum\limits_{m = 0}^{{K_1} - 1} {}\sum\limits_{l = 0}^{{L_{\mathrm{FIR1}}} - 1} {}h_{1,m}^*[l]r[n - i - l]{e^{ - j2\pi {\alpha _m}\left( {n - i} \right)}}   \Big){e^{ - j2\pi {\beta _k}n}}$. Let us define $\tilde h_{k,m}[i,l] \triangleq h_{2,k}[i]\big(\delta [{\alpha _m}]\delta [l] - h_{1,m}[l]\big)$, we can now write
\vspace{-0.2cm}
\begin{equation*}
y[n]=\sum\limits_{k = 0}^{{K_2} - 1} {}\sum\limits_{i = 0}^{{L_{\mathrm{FIR2}}} - 1} {}\sum\limits_{m = 0}^{{K_1} - 1} {}\sum\limits_{l = 0}^{{L_{\mathrm{FIR1}}} - 1} {}\tilde h_{k,m}^*[i,l]r[n - i - l]  {e^{ - j2\pi \left( {{\alpha _m}(n-i) + {\beta _k}n} \right)}}.
\vspace{-0.2cm}
\end{equation*}
For a received signal of the form $r[n]=d[n]+w[n]$, where $d[n]$ and $w[n]$ are mutually independent and the noise has a zero mean, we have
\vspace{-0.2cm}
\begin{equation}
\label{eqn:scale_11}
\E\left\{ {y[n]|d[n]} \right\} = \sum\limits_{k = 0}^{{K_2} - 1} {\sum\limits_{i = 0}^{{L_{\mathrm{FIR2}}} - 1} {}\sum\limits_{m = 0}^{{K_1} - 1} {}\sum\limits_{l = 0}^{{L_{\mathrm{FIR1}}} - 1} {}\tilde h_{k,m}^*[i,l]\E\big\{ {d[n - i - l]|d[n]} \big\}  {e^{ - j2\pi \left( {{\alpha _m}(n-i) + {\beta _k}n} \right)}}}.
\vspace{-0.2cm}
\end{equation}

Recall that the desired signal is an OFDM signal with a symbol period of $N_{sym}$ samples, where each symbol contains $N_{cp}$ cyclic prefix samples followed by $N_{data}=N_{sym}-N_{cp}$ data samples.
Assuming that the number of subcarriers is large enough and the data symbols are i.i.d, then, from the central limit theorem (CLT), it follows that, the PDF of each time-domain sample of the OFDM signal converges to a Gaussian distribution, and that the PDF of each baseband sample converges to a proper complex Gaussian distribution, see details in~\cite[Pg. 120]{Nee:00} and~\cite{Ochiai:01}.
We now show that the time-domain samples are in-fact jointly Gaussian. Note that if the samples are taken from different OFDM symbols they are independent and therefore jointly Gaussian. For samples that are both taken from the $m$-th symbol, we define
\vspace{-0.2cm}
\begin{equation*}
{\bf{u}}[n,i] \triangleq  \left[ {\begin{array}{*{20}{c}}
{d[n]}\\
{d[n - i]}
\end{array}} \right] = \frac{1}{{\sqrt {{N_{data}}} }}\sum\limits_{k = 0}^{{N_{data}} - 1} {{{\bf{u}}_k}[n,i]} ,
\vspace{-0.2cm}
\end{equation*}
and
\vspace{-0.2cm}
\begin{equation*}
{{\bf{u}}_k}[n,i] \triangleq \left[ {\begin{array}{*{20}{c}}
{\left| {{a_{m,k}}} \right|\cos \left( {\left( {\phi  + {\omega _k}} \right)n - {\omega _k}m' + {\varphi _{m,k}}} \right)}\\
{\left| {{a_{m,k}}} \right|\cos \left( {\left( {\phi  + {\omega _k}} \right)\left( {n - i} \right) - {\omega _k}m' + {\varphi _{m,k}}} \right)}
\end{array}} \right].
\vspace{-0.2cm}
\end{equation*}
where $\phi  \buildrel \Delta \over = 2\pi {f_c}{T_{samp}}$, $m' = m{N_{sym}}$, ${\omega _k} = \frac{{2\pi k}}{{{N_{data}}}}$,
and ${a_{m,k}} = \left| {{a_{m,k}}} \right|{e^{j{\varphi _{m,k}}}}$.
It is simple to verify that the vectors ${{\bf{u}}_k}[n,i]$ satisfy the multivariate Lindeberg-Feller conditions~\cite[Pg. 913]{Greene:03}, and that the PDF of ${\bf{u}}[n,i]$ converges to a multivariate Gaussian distribution, therefore $d[n]$ and $d[n-i]$ are jointly Gaussian.
 Note that  $\E\left\{ {d[n - i]|d[n]} \right\} = d[n]$ when $i=0$. Next, define  $\gamma \triangleq \phi{N_{data}}$. When $d[n - i]$ is a cyclic prefix replica of $d[n]$, i.e., $i=N_{data}$, we obtain $\E\left\{ {d\left[ {n - {N_{data}}} \right]|d\left[ n \right]} \right\} = d\left[ n \right]\cos (\gamma ) - \frac{1}{2}\E\Big\{ {\mathop{\rm Im}\nolimits} \big\{ {s\left[ n \right]{e^{ - j\phi n}}} \big\}\Big|{\mathop{\rm Re}\nolimits} \big\{ {s\left[ n \right]{e^{ - j\phi n}}} \big\} \Big\}\sin ( \gamma ) $. Since $ s[n]$ is a zero mean proper complex Gaussian RV, we obtain $\E\left\{ {d\left[ {n - {N_{data}}} \right]|d\left[ n \right]} \right\} = d\left[ n \right]\cos ( \gamma  )$.
For other values of $i$, it follows from \eqref{eqn:OFDM_autocorr4} that the time-domain OFDM samples $d[n]$ and $d[n-i]$ are uncorrelated. As $\big(d[n], d[n-i]\big)$ are jointly Gaussian, then they are statistically independent, that is  $\E\left\{ {d[n - i]|d[n]} \right\} = \E\left\{ {d[n - i]} \right\} = 0$. From the above discussion it follows that
$\E\left\{ {d[n - i]|d[n]} \right\} = d[n]\kappa[n,i]$, where $\kappa[n,i] \triangleq \big({\bf{1}}_{\mathcal{S}_{CP}}[n]\delta[i-N_{data}]\cos(\gamma ) + \delta[i]\big)$.

By plugging the expression for $\E\left\{ {d[n - i]|d[n]} \right\}$ into \eqref{eqn:scale_11}, the scaling at a given index $n$ is \\
\vspace{-0.2cm}
\begin{equation}
\label{eqn:scale_14}
\psi [n] = \sum\limits_{k = 0}^{{K_2} - 1} \sum\limits_{i = 0}^{{L_{\mathrm{FIR2}}} - 1} \sum\limits_{m = 0}^{{K_1} - 1} \sum\limits_{l = 0}^{{L_{\mathrm{FIR1}}} - 1}  {\tilde h_{k,m}^*[i,l]\kappa[n,i+l]} 
{e^{ - j2\pi \left( {{\alpha _m}(n-i) + {\beta _k}n} \right)}}   .
\vspace{-0.2cm}
\end{equation}

\vspace{-0.4cm}
\subsection{Application of the New Algorithm to ISI Channels}
\label{subsec:Multipath}
\vspace{-0.25cm}
The algorithm proposed in Subsection \ref{subsec:Noise_Estimate} assumes the received signal is in the form of $r[n]=d[n]+w[n]$ where $d[n]$ denotes the desired time-domain OFDM signal and $w[n]$ denotes the cyclostationary PLC noise. In this section we will show that, assuming the receiver knows the channel, then ISI can be easily incorporated into the proposed model.

Consider the OFDM signal received over an ISI channel given by
$r[n] = \sum\limits_{i = 0}^{{L_{ISI}} - 1} {g[i]d[n - i]}  + w[n]$,
where $g[i]$, $i\in\big\{0,1,\ldots,L_{ISI}-1\big\}$, are channel coefficients, $d[n]$ is the time-domain OFDM signal with a symbol length of $N_{sym}$ samples, see \eqref{eqn:OFDM1}, and $w[n]$ is the additive cyclostationary noise.
The received signal may therefore be written as $r[n]=d_{ISI}[n]+w[n]$ where the desired signal component is obtained by
${d_{ISI}}[n] = \sum\limits_{i = 0}^{{L_{ISI}} - 1} {g[i]d[n - i]}$.
Note that $d[n]$ is a time-domain OFDM signal, therefore, as shown in Subsection~\ref{subsec:cyclo_OFDM}, it is a zero-mean cyclostationary stochastic process with period of $N_{sym}$. Hence, the mean value of $d_{ISI}[n]$ is obtained by
$\E\left\{ {{d_{ISI}}[n]} \right\} = \E\left\{ {\sum\limits_{i = 0}^{{L_{ISI}} - 1} {g[i]d[n - i]} } \right\} = \sum\limits_{i = 0}^{{L_{ISI}} - 1} {g[i]\E\left\{ {d[n - i]} \right\}}  = 0$,
and the autocorrelation function of $d_{ISI}[n]$ is obtained by
${c_{{d_{ISI}}{d_{ISI}}}}\left( {n,l} \right) = \E\left\{ {{d_{ISI}}[n + l]d_{ISI}^*[n]} \right\}= \sum\limits_{i = 0}^{{L_{ISI}} - 1} {\sum\limits_{k = 0}^{{L_{ISI}} - 1} {g[i]{g^*}[k]} } {c_{dd}}\left( {n - k,l + k - i} \right)$.
Observe that $\E\left\{ {{d_{ISI}}[n]} \right\} = \E\left\{ {{d_{ISI}}[n+N_{sym}]} \right\}$ and that ${c_{{d_{ISI}}{d_{ISI}}}}\left( {n + {N_{sym}},l} \right) = {c_{{d_{ISI}}{d_{ISI}}}}\left( {n,l} \right)$. Thus, the desired signal component $d_{ISI}[n]$ is cyclostationary with the same period as the OFDM signal.
We conclude that the design of the FRESH filter for recovery of an OFDM signal, received via an ISI channel with an additive cyclostationary noise, is done in the following two steps:
\begin{itemize}
	\item {Design the FRESH filter to recover $d_{ISI}[n]$ from the received signal $r[n]$.}
	\item {Decode the data symbols from the recovered $d_{ISI}[n]$. As inter-symbol interference is inherently handled by the OFDM signal detection process, the FRESH filter is not designed to remove the ISI, but to recover $d_{ISI}[n]$.}
\end{itemize}
Note that the cyclostationarity is also maintained in case the channel is LPTV (e.g., $g[i]$ is replaced with  $g[n,i]$, where for some integer $N_{ch}$, $g[n,i] = g[n + N_{ch},i]$). Recall that the filtering of a cyclostationary signal by an LPTV system results in a cyclostationary signal
~\cite[Sec. 17.4.4]{Giannakis:98}. Thus, our receiver structure is the same for LPTV channels if TA-MSE is the design criterion.

\vspace{-0.4cm}
\subsection{Comparison with Existing Solutions}
\label{subsec:Existing_Solutions}
\vspace{-0.25cm}
A variety of methods exists for exploiting the time-domain cyclic redundancy of an OFDM signal, induced by the cyclic prefix, for various estimation tasks.
The work in~\cite{Tarighat:03} improves decoding performance by combining the received time-domain OFDM samples with their corresponding CP samples, prior to discarding the CP samples and decoding the OFDM signal. The CP combining is implemented via a sub-optimal least-squares algorithm realized by averaging the received samples with their corresponding cyclic replicas. Note that, among all time-domain pre-combining methods, the FRESH filter results in the minimal TA-MSE, as it is derived analytically from the multivariate Wiener filtering problem~\cite{Gardner:93}. In addition, our proposed receiver algorithm is independent of the OFDM decoder and may therefore be combined with any OFDM decoding algorithm. Thus, our algorithm has a lower TA-MSE than the algorithm proposed in~\cite{Tarighat:03}. 

Another class of schemes uses the cyclostationary nature of the narrowband PLC noise for improving performance. An interesting recent work in~\cite{Evans:12} proposed a linear periodic time-varying filter to whiten the Gaussian cyclostationary noise.  We note that the frequency-domain representation of cyclostationary processes is obtained by the {\em two-dimensional} cyclic spectra~\cite{Giannakis:98, Gardner:06}, rather than by the one-dimensional Fourier transform. Therefore, whitening the one-dimensional Fourier transform of the noise by weighting the frequency bins with periodically time-varying weights does not fully exploit the redundancy present in the cyclostationary  noise, and may even deteriorate performance at low SNR. This is in contrast to our proposed algorithm which is beneficial also at low SNR.

\vspace{-0.4cm}
\subsection{Complexity Analysis}
\label{subsec:complex}
\vspace{-0.25cm}
In this section the complexity of the receiver proposed in Subsection~\ref{subsec:Noise_Estimate} is analyzed and compared with other algorithms. Note that the algorithm consists of two FRESH filters in series: $h_1[n]$ which consists of $K_1$ branches, each includes an LTI filter with $L_{\mathrm{FIR1}}$ taps, and $h_2[n]$ which consists of $K_2$ branches, each includes an LTI filter with $L_{\mathrm{FIR2}}$ taps. The complexity of the receiver may therefore be measured by a total of $K_1\times L_{\mathrm{FIR1}} + K_2\times L_{\mathrm{FIR2}}$ taps.
Recall that $K_1$ and $K_2$ are upper-bounded by $N_{sym}$ and $N_{noise}$, respectively. In practice, as shown in~\cite{Chen:11}, the number of branches may be much smaller, as long as the branches are symmetric around cyclic frequency zero and a branch which corresponds to cyclic frequency zero is included.
In order to exploit the cyclostationarity induced by the cyclic prefix, $L_{\mathrm{FIR1}}$ must be at least $N_{data}$. From Subsection \ref{subsec:cyclo_noise} we note that for large enough values of $l$, $c_{ww}(n,l) \approx 0$ for all $n$, therefore the value of $L_{\mathrm{FIR2}}$ should be sufficiently large so that $c_{ww}(n,l) \approx 0$ for all $l \geq L_{\mathrm{FIR2}}$  .

We now compare the complexity of our proposed algorithm to the algorithm proposed in~\cite{Evans:12}. The work in~\cite{Evans:12} assumes the LPTV noise model proposed in~\cite{Nassar:12}, see details in Subsection \ref{subsec:cyclo_noise}.
The receiver proposed in~\cite{Evans:12} filters the received signal by $M$ LTI whitening filters in parallel, and at each interval the output of the receiver is obtained as the output of the corresponding filter. As the whitening LTI filters apply a different weight to each frequency bin, the number of taps at each filter must be at least the number of frequency bins, that is $N_{data}$. The complexity of the algorithm proposed in~\cite{Evans:12} may therefore be evaluated by a total of $M\times N_{data}$ taps.
We conclude that the complexity of our proposed algorithm is of the same scale as that of the only other algorithm designed for narrowband PLC performance improvement.

\vspace{-0.25cm}
\section{An Adaptive Implementation of the Proposed Algorithm}
\label{sec:Adaptive}
\vspace{-0.25cm}
In order to derive the optimal FRESH filter, the autocorrelation of the frequency shifted input vector and the cross-correlation between the input vector and the desired signal are required. In practical systems, it is unlikely that these values are known a-priori, therefore, an adaptive implementation of the FRESH filter is necessary in order to incorporate such filtering in practice.

The adaptive FRESH filter is schematically depicted in Fig. \ref{fig:AdaptiveModel2}:
\begin{figure}
	\centering
		\includegraphics[width=0.5\textwidth]{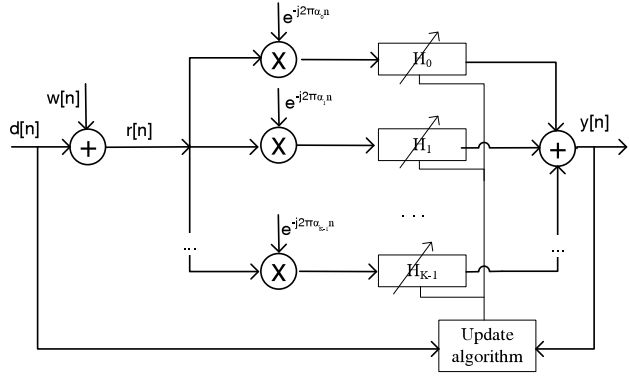}
\vspace{-0.4cm}
	\caption{Adaptive FRESH filter system model with ideal training signal.}
	\label{fig:AdaptiveModel2}
	\vspace{-0.8cm}
\end{figure}
The output of the FRESH filter, as well as an ideal reference signal, are provided as inputs to an adaptive algorithm, which updates the filter coefficients according to the error between the two signals.
In practical systems, a reference signal may be obtained from two sources
\begin{itemize}
	\item {A training signal: When the data is a-priori known to the receiver (as in Fig. \ref{fig:AdaptiveModel2}).}
	\item {A decision directed reference signal: Using the decisions provided by the receiver as training for the adaptive algorithm.}
\end{itemize}
It is possible to combine both methods by including a preamble followed by a data signal at every OFDM frame (which may include up to thousands of OFDM symbols): the preamble is used as a training signal during preamble transmission, and a decision directed reference signal is used for tracking during data transmission.

\vspace{-0.4cm}
\subsection{Initial FRESH Filter Acquisition: Exponential RLS Adaptive Algorithm Based on Training}
\label{sec:ExpRLS}
\vspace{-0.25cm}
The exponential recursive least squares (RLS) algorithm~\cite{Haykin:03} is an adaptive algorithm which minimizes the cost function
$\varepsilon [n] = \sum\limits_{i = 1}^n {{\lambda ^{n - i}}{{\left| {e[i]} \right|}^2}} $,
where $e[n] = d[n] - y[n] = d[n] - {{\bf{h}}^H}[n]{\bf{z}}[n]$, and $0 < \lambda \leq 1$ represents the memory of the algorithm.
As shown in~\cite[Ch. 13]{Haykin:03}, the filter ${\bf{h}}[n]$ which minimizes $\varepsilon [n]$ is the solution of the equation $\left( {\sum\limits_{i = 1}^n {{\lambda ^{n - i}}{\bf{z}}[i]{{\bf{z}}^H}[i]} } \right) {\bf{h}}[n] = \sum\limits_{i = 1}^n {{\lambda ^{n - i}}{\bf{z}}[i]{d^*}[i]}$.

The adaptive algorithm updates during runtime the vector $ {\bf{h}}[n]$, which is initialized to ${\bf{h}}[0] = {{\bf{i}}_{{L_{\mathrm{FIR}}} \cdot {k_0}}} $, and a matrix ${\bf{P}}[n]$ approximating the value of ${\left( {\sum\limits_{i = 1}^n {{\lambda ^{n - i}}{\bf{z}}[i]{{\bf{z}}^H}[i]} } \right)^{ - 1}}$, which is initialized to ${\bf{P}}[0] = \epsilon {{\bf{I}}_{{L_{FIR}} \cdot K}}$ where $\epsilon$ is a small positive constant and ${\bf I}_N$ denotes the $N\times N$ identity matrix. At each time instant, the algorithm executes the following computations:
\begin{enumerate}
	\item {Compute an estimate of the a-priori error via $\xi [n] = d[n] - {{\bf{h}}^H}[n - 1]{\bf{z}}[n]$.}
	\item {Compute a gain vector as ${\bf{k}}[n] = \frac{{{\lambda ^{ - 1}}{\bf{P}}[n - 1]{\bf{z}}[n]}}{{1 + {\lambda ^{ - 1}}{{\bf{z}}^H}[n]{\bf{P}}[n - 1]{\bf{z}}[n]}}$.}
	\item {Update the matrix ${\bf{P}}[n]$ according to $		{\bf{P}}[n] = {\lambda ^{ - 1}}{\bf{P}}[n - 1] - {\lambda ^{ - 1}}{\bf{k}}[n]{{\bf{z}}^H}[n]{\bf{P}}[n - 1]$.}
	\item {Update the vector ${\bf{h}}[n]$ according to ${\bf{h}}[n] = {\bf{h}}[n - 1] + {\bf{k}}[n]{\xi ^*}[n]$.}		
\end{enumerate}
When the training signal is equal to the desired signal without errors, it was shown in~\cite[Ch. 13]{Haykin:03} that the RLS algorithm always obtains the minimal value of the cost function, and ${\bf{h}}[n]$ converges to the optimal FRESH filter in the ergodic sense.

\vspace{-0.4cm}
\subsection{Tracking Phase: Decision Directed Adaptive FRESH}
\label{sec:DDFresh}
\vspace{-0.25cm}
As noted in the previous subsection, during transmission of information the filter coefficients must be updated using the decision directed approach, namely the decoded bit stream at the output of the receiver is used to generate the training signal.
This situation is demonstrated in Fig. \ref{fig:DecisionDirect1}:
\begin{figure}
	\centering
		\includegraphics[width=0.5\textwidth]{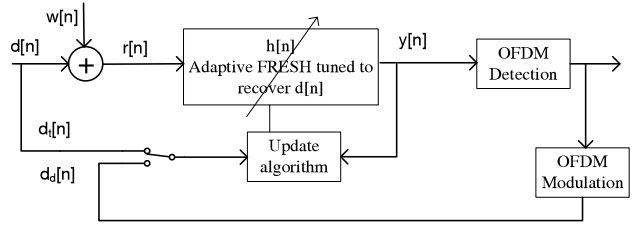}
\vspace{-0.4cm}
	\caption{Decision directed adaptive FRESH filter.}
	\label{fig:DecisionDirect1}
	\vspace{-0.8cm}
\end{figure}
the filtered signal $y[n]$ is fed into the OFDM detector block which generates the decoded bit stream. This bit stream is encoded and modulated to produce the decision-based reference OFDM signal $d_d[n]$. Now, during the preamble transmission (i.e., initial acquisition), the adaptive algorithm uses the a-priori known preamble signal $d_t[n]$ as a reference signal. Then, when the new data is transmitted, the adaptive algorithm uses the decision-based reference signal $d_d[n]$ to evaluate the error and update the filter coefficients.

\vspace{-0.4cm}
\subsection{Decision Directed Adaptive Signal Recovery with Noise Estimation}
\label{sec:DDFresh_wNC}
\begin{figure}
	\centering
		\includegraphics[width=0.75\textwidth]{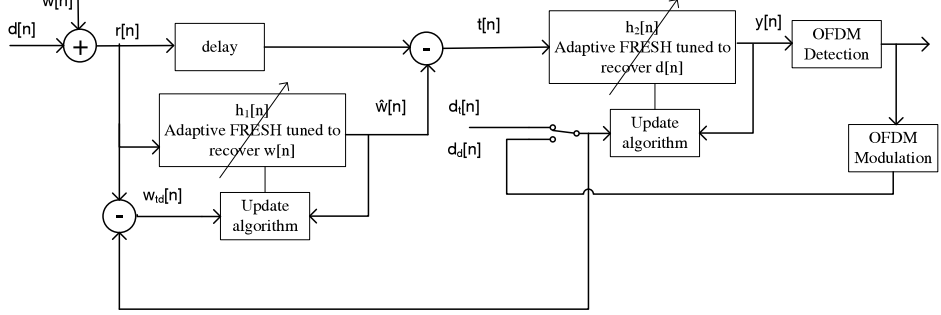}
\vspace{-0.4cm}
	\caption{Decision directed adaptive receiver with noise cancellation.}
	\label{fig:DecisionDirect2}
	\vspace{-1.0cm}
\end{figure}

\vspace{-0.25cm}
We now describe the adaptive implementation of the overall scheme with noise cancellation proposed in Subsection \ref{subsec:Noise_Estimate}. This structure is depicted in Fig. \ref{fig:DecisionDirect2}.
Since the proposed receiver includes a FRESH filter tuned to recover the cyclostationary noise $w[n]$, it requires a reference signal to adapt the noise estimation filter $h_1[n]$, which is done as described in Subsection \ref{sec:DDFresh}. This signal is obtained by subtracting from the received signal $r[n]$ the reference signal for the FRESH filter $h_2[n]$, tuned to recover the OFDM signal $d[n]$.
The adaptive FRESH filter $h_2[n]$, is identical to the implementation detailed in Subsection \ref{sec:DDFresh}.
Note that the adaptive receiver with noise estimation updates its coefficients at each detected {\em codeword} rather than at each incoming {\em sample}, contrary to the standard implementation of the RLS algorithm which can be found in \cite{Haykin:03}.
It should be also noted that the new receiver proposed in this section is likely to be more sensitive to detection errors than a receiver that consists of a single FRESH filter, as each such error affects the coefficients of both $h_1[n]$ and $h_2[n]$.

The adaptive implementation requires the receiver to know only the periods of the cyclostationary noise and of the OFDM signal, denoted $N_{noise}$ and $N_{sym}$, respectively. For practical PLC scenarios these values are a-priori known: $N_{sym}$ is known by design and the period of the noise, $N_{noise}$, is known to be half the AC cycle (see \cite{Pavlidou:03, Hooijen:98, Zimmermann:02a, Zimmermann:02b, Katayama:06, Nassar:12}). Thus, the adaptive implementation is very robust to noise model parameters.
The a-priori fixed implementation, on the other hand, requires the receiver to know the autocorrelation functions of the cyclostationary noise and of the OFDM signal, denoted $c_{ww}(n,l)$ and $c_{dd}(n,l)$, respectively, and is therefore more susceptible to noise model parameters.
Lastly, we note that if $N_{noise}$ is unknown, the performance of the receiver derived in this work converges to that of the direct signal recovery receiver of~\cite{Chen:11}.

\vspace{-0.25cm}
\section{Simulations}
\label{sec:Simulations}
\vspace{-0.25cm}
In this section, the performance of the receiver developed in Section \ref{sec:Model}, as well as that of the adaptive implementation developed in Section \ref{sec:Adaptive}, are evaluated by simulations, and compared with the algorithm proposed in~\cite{Chen:11}.
The information bits are encoded in accordance with the IEEE P1901.2 standard~\cite{IEEE:13}:  an outer Reed-Solomon $(255,239)$ code is followed by an inner rate $\frac{1}{2}$ convolutional code with generator polynomials $171_{octal}$ and $155_{octal}$, and an interleaver specified in~\cite{IEEE:13}.
The information signal is a passband OFDM signal with $32$ subcarriers over the frequency band $3 - 148.5$ kHz, each modulated with QPSK constellation. This frequency range is in accordance with the European CENELEC regulations~\cite{CA:91}. We use a CP consisting of $16$ samples, hence the total number of samples at each OFDM symbol is $80$.

Three types of noise are simulated -
\begin{enumerate}
\item {ACGN based on the LPTV model~\cite{Nassar:12} adopted by the IEEE P1901.2 standard~\cite{IEEE:13}, with two sets of typical parameters, referred to in the following as ${\mbox{IEEE1}}$ and ${\mbox{IEEE2}}$:
\begin{itemize}
	\item {${\mbox{IEEE1}}$ corresponds to low voltage site 8 (LV8) in~\cite[Appendix G]{IEEE:13}.}
	\item {${\mbox{IEEE2}}$ corresponds to low voltage site 14 (LV14) in~\cite[Appendix G]{IEEE:13}.}
\end{itemize}
}
\item {ACGN based on the Katayama model~\cite{Katayama:06} with two sets of typical parameters, referred to in the following as ${\mbox{KATA1}}$  and ${\mbox{KATA2}}$:
\begin{itemize}
	\item {The parameters for ${\mbox{KATA1}}$ are taken from~\cite{Katayama:06} and are set to be $\left\{ {n_0}, {n_1}, {n_2} \right\} = \{ 0,1.91,1.57 \cdot {10^5}\} $, $\left\{ {{\Theta _0}}, {\Theta _1}, {\Theta _2} \right\} = \{ 0, - 6,$ $- 35\} $ degrees, $\left\{ {{A_0}}, {A_1}, {A_2} \right\} = \{ 0.23, 1.38,$ $7.17\} $, and $\alpha_1 = 1.2\cdot 10^{-5}$.}
	\item {The parameters for ${\mbox{KATA2}}$ are taken from~\cite[residence 1]{Katayama:00}, and are set to be $\left\{ {n_0}, {n_1}, {n_2} \right\} = \{ 0,9.3,5.3 \cdot {10^3}\} $, $\left\{ {{\Theta _0}}, {\Theta _1}, {\Theta _2} \right\} = \{ 0, 128, 161\} $ degrees, $\left\{ {{A_0}}, {A_1}, {A_2} \right\} = \{ 0.13, 2.8, 16\} $, and $\alpha_1 = 8.9\cdot 10^{-6}$.}
\end{itemize}
}
\item {AWGN (in order to show robustness to the noise model).}
\end{enumerate}
Note that ${\mbox{IEEE1}}$, ${\mbox{IEEE2}}$, and ${\mbox{KATA2}}$ correspond to practical periodic impulsive component with duration of $300-400$ microseconds, while ${\mbox{KATA1}}$ corresponds to a periodic impulsive component with a very short duration of $25$ microseconds, representing very unfavorable conditions.
The cyclic period of the cyclostationary noise is set to $N_{noise} = 1000$ samples.
Note that the noise period, $N_{noise}$, and the length of the OFDM symbol, $N_{sym}$, are both scaled by a factor of $\frac{1}{2.5}$ compared to their practical values to reduce simulation time. However, as $\frac{N_{sym}}{N_{noise}}$ is the same as in practical systems, the results correspond to the performance of practical systems.

Four receivers are simulated -
\begin{enumerate}
\item {${\mbox{Rx}}_1$: A receiver with {\em no filtering} applied to the input signal prior to decoding.}
\item {${\mbox{Rx}}_2$: A receiver which implements a {\em stationary FIR Wiener filter}~\cite[Ch. 12.7]{Kay:93} with $N_{sym} + \frac{N_{noise}}{2}$ taps applied to the input signal $r[n]$.}
\item {${\mbox{Rx}}_3$: {\em Best of previous work} is represented by a receiver with a FRESH filter tuned to extract the desired OFDM signal~\cite{Chen:11}. The filter utilizes 5 cyclic frequencies in the range  $\frac{-2}{N_{sym}},\ldots,\frac{2}{N_{sym}}$ such that each FIR has $N_{sym} + \frac{N_{noise}}{2}$ taps.}
\item {${\mbox{Rx}}_4$: {\em Our newly proposed algorithm} is demonstrated by a receiver with the FRESH filter $h_1[n]$ utilizing 5 cyclic frequencies in the range  $\frac{-2}{N_{noise}},\ldots,\frac{2}{N_{noise}}$, and at each branch the FIR has ${L_{\mathrm{FIR1}}}=\frac{N_{noise}}{2}$ taps; and the FRESH filter $h_2[n]$ utilizing 5 cyclic frequencies in the range $\frac{-2}{N_{sym}},\ldots,\frac{2}{N_{sym}} $, and at each branch the FIR has ${L_{\mathrm{FIR2}}}=N_{sym}$ taps.}
\end{enumerate}
Note that ${\mbox{Rx}}_2$, ${\mbox{Rx}}_3$ and ${\mbox{Rx}}_4$ have the same delay, and that the new receiver (${\mbox{Rx}}_4$) and the scheme of \cite{Chen:11} (${\mbox{Rx}}_3$) have the same number of coefficients. We also note that the stationary Wiener filter (${\mbox{Rx}}_2$) has less coefficients than our new receiver ${\mbox{Rx}}_4$, but it has the same delay. Increasing the number of taps in ${\mbox{Rx}}_2$ increases the delay but does not improve the performance of ${\mbox{Rx}}_2$ in the simulations.
The results are plotted for various values of input SNR defined as ${\mbox{SNR}}_{in} \triangleq \frac{P_d}{\left\langle \E\left\{w[n]w^*[n]\right\} \right\rangle}$.
For evaluating the bit error rate (BER) performance, a per-subcarrier maximum likelihood (ML) decoder (i.e., minimal Euclidean distance decoder) is used.

\vspace{-0.5cm}
\subsection{Simulation Study of the Optimal Receiver}
\label{sec:Sim_Optimal}
\vspace{-0.25cm}
In this section the performance of the receiver developed in Section \ref{sec:Model} is evaluated. Four aspects were studied: first, a comparison between the simulation results and the theoretical results was done to confirm the validity of the simulations. Then, the TA-MSE performance was evaluated and the robustness of the new receiver ${\mbox{Rx}}_4$ to the noise model was tested. Next, the BER and the corresponding input SNR gains were evaluated, and lastly, the applicability of the new receiver with noise cancellation to multipath channels was demonstrated.

\begin{figure}
\centering
\begin{minipage}{.45\linewidth}
  \includegraphics[width=3in,clip=true, viewport=0.75in 0.2in 7.6in 5.6in]{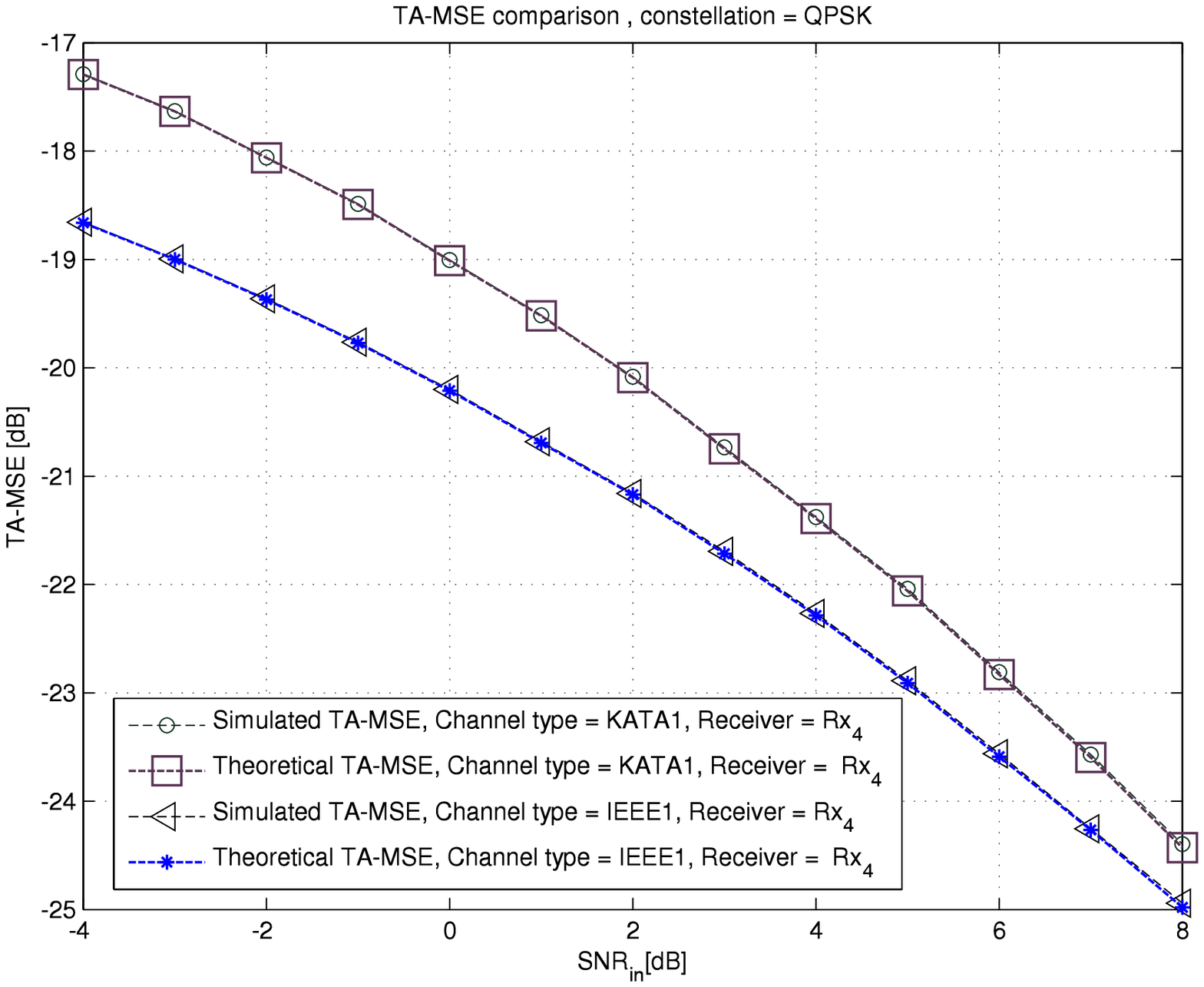}
  \vspace{-0.8cm}
  \caption{Theoretical TA-MSE comparison with simulated TA-MSE.}
  \label{fig:SNRTheo_Compare}
\end{minipage}
\quad
\begin{minipage}{.45\linewidth}
  \includegraphics[width=3in,clip=true, viewport=0.75in 0.2in 7.6in 5.6in]{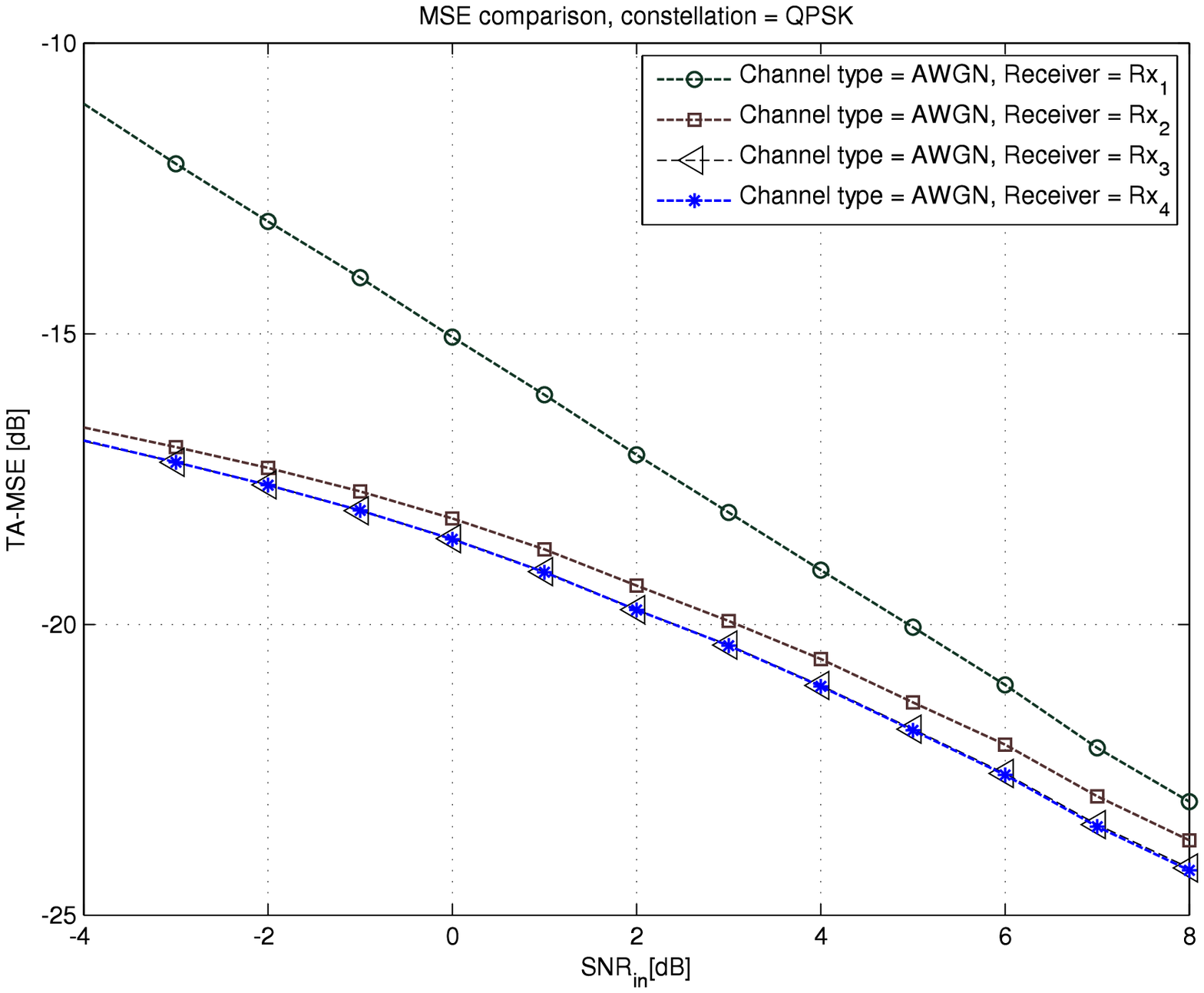}
  \vspace{-0.8cm}
  \caption{TA-MSE comparison for AWGN channel to verify robustness to noise model.}
  \label{fig:SNR_Out_Compare_AWGN}
\end{minipage}
\vspace{-0.2cm} 
\end{figure}

{ \bf{\em{1) Verifying Agreement Between Analytical Results and Simulation}:}}
We first verified the agreement between the simulation and analytical TA-MSE expression \eqref{eqn:model2_output2}. To that aim we simulated both noise models; For the Katayama model~\cite{Katayama:06} we used the parameters set ${\mbox{KATA1}}$, and for the LPTV model~\cite{Nassar:12} we used the parameters set ${\mbox{IEEE1}}$.
In Fig. \ref{fig:SNRTheo_Compare} both the analytical TA-MSE evaluated using \eqref{eqn:model2_output2} and the TA-MSE evaluated from the simulation output are compared. We observe that there is {\em an excellent agreement between the analytical and simulated TA-MSE for both noise models}. This confirms the validity of our simulation study described next.

{ \bf{\em{2) Evaluating TA-MSE Performance and Verifying Robustness of the New Algorithm to the Noise Model}:}}
We next evaluated the TA-MSE performance and tested the robustness of the proposed receiver algorithm to the exact noise model. First, we verified that the new receiver operates well also in AWGN. The simulation results are presented in Fig. \ref{fig:SNR_Out_Compare_AWGN}. Observe that when the new receiver ${\mbox{Rx}}_4$ is applied to AWGN it achieves the same TA-MSE performance as the optimal FRESH filter without noise cancellation of~\cite{Chen:11} (${\mbox{Rx}}_3$), which is tuned to recover only the OFDM signal.
This shows the robustness of the new receiver to the noise model, as the noise cancellation part in ${\mbox{Rx}}_4$, {\em which cannot provide improvement in AWGN}, does not degrade performance of ${\mbox{Rx}}_4$. Both ${\mbox{Rx}}_4$ and ${\mbox{Rx}}_3$ achieve $0.8$ dB input SNR gain over the stationary Wiener filter (${\mbox{Rx}}_2$) for ${\mbox{SNR}}_{in}\leq 2$ dB, and the gain decreases to $0.55$ dB at ${\mbox{SNR}}_{in}=6$ dB.

\begin{figure}
\centering
\begin{minipage}{.45\linewidth}
  \includegraphics[width=3in,clip=true, viewport=0.75in 0.2in 7.6in 5.6in]{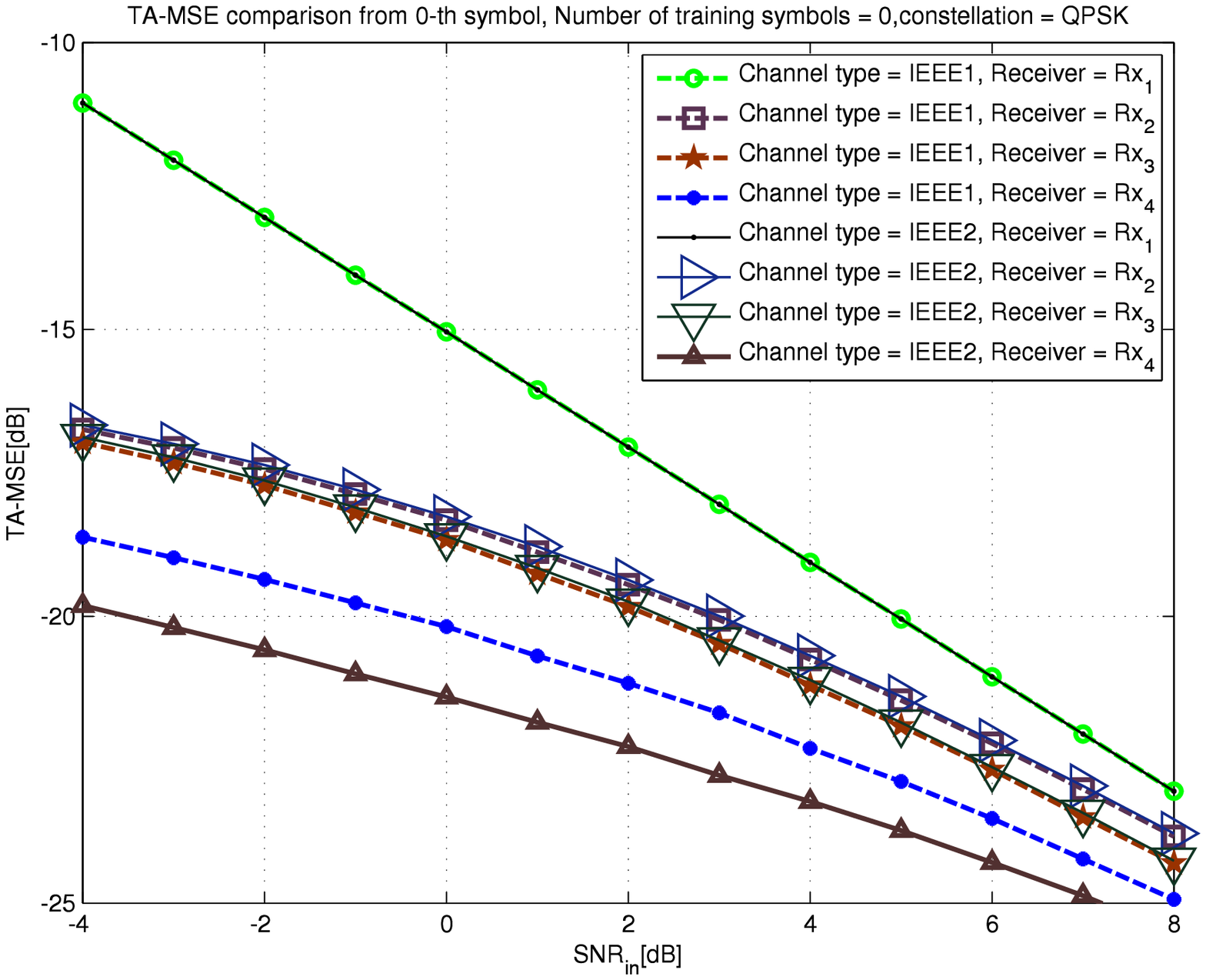}
  \vspace{-0.8cm}
  \caption{TA-MSE comparison for the LPTV noise model of~\cite{Nassar:12}.}
  \label{fig:SNR_Out_Compare_LPTV}
\end{minipage}
\quad
\begin{minipage}{.45\linewidth}
  \includegraphics[width=3in,clip=true, viewport=0.75in 0.2in 7.6in 5.6in]{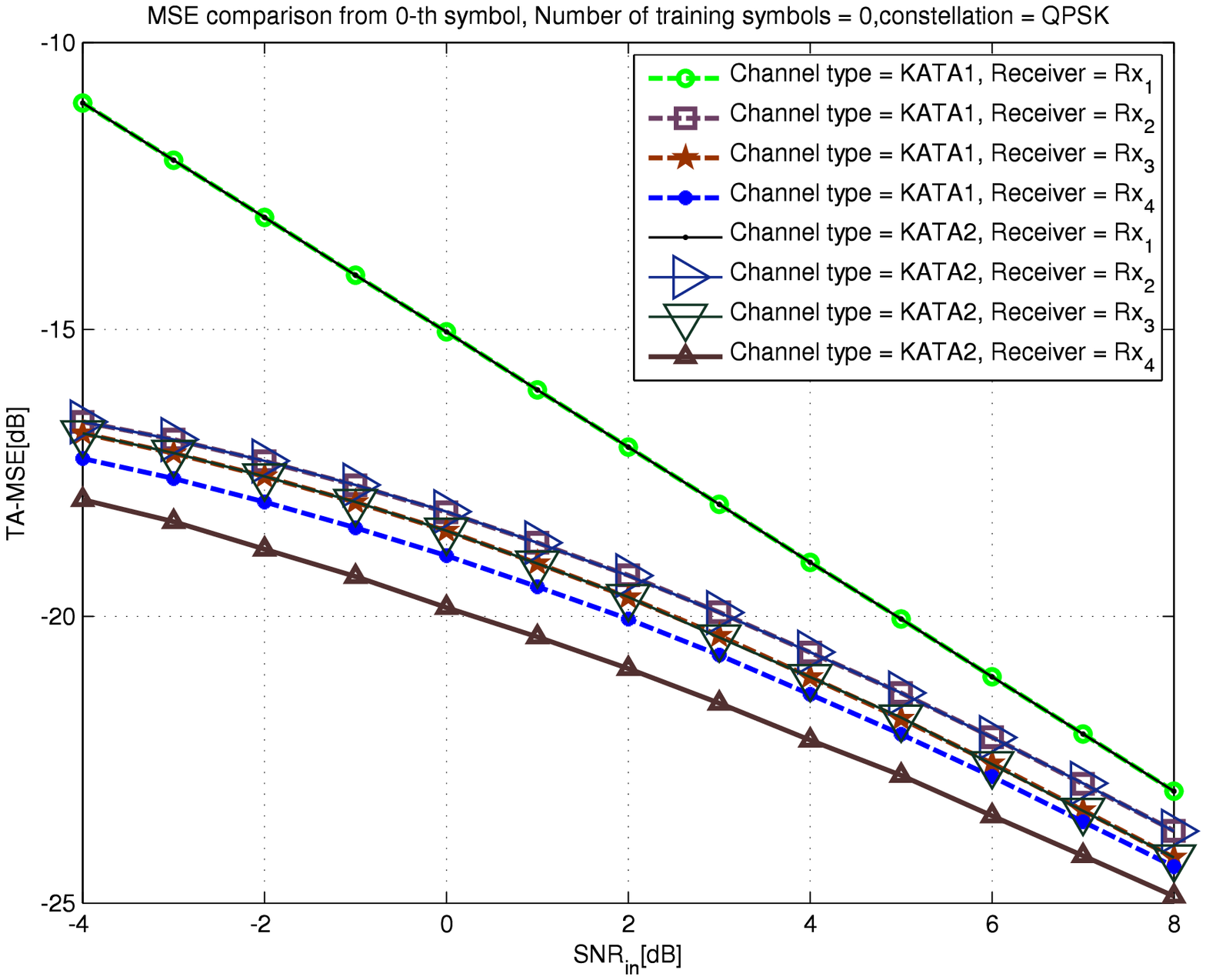}
  \vspace{-0.8cm}
  \caption{TA-MSE comparison for the Katayama noise model of~\cite{Katayama:06}.}
  \label{fig:SNR_Out_Compare_Cycsta}
\end{minipage}
\vspace{-1.0cm} 
\end{figure}

Next, the TA-MSE was evaluated for four sets of ACGN models:
The results for the LPTV model~\cite{Nassar:12} with parameters ${\mbox{IEEE1}}$ and ${\mbox{IEEE2}}$ are depicted in Fig. \ref{fig:SNR_Out_Compare_LPTV}.
The results for the Katayama model~\cite{Katayama:06} with parameters ${\mbox{KATA1}}$ and ${\mbox{KATA2}}$ are depicted in Fig. \ref{fig:SNR_Out_Compare_Cycsta}.
Observe that the performance improvement depends on the cyclostationary characteristics of the noise:
When the impulsive noise is of typical width of $300-400$ microseconds as in ${\mbox{IEEE1}}$, ${\mbox{IEEE2}}$ and ${\mbox{KATA2}}$, the noise has a stronger cyclic redundancy and therefore noise cancellation is more effective.
Accordingly, for the ${\mbox{IEEE}}$ models we observe in Fig. \ref{fig:SNR_Out_Compare_LPTV} input SNR gains of $2.5-6$ dB compared to ${\mbox{Rx}}_3$ at ${\mbox{SNR}}_{in} \leq 0$ dB, which decreases at ${\mbox{SNR}}_{in}=4$ dB to a $2.7$ dB gain for ${\mbox{IEEE2}}$ and a $1.55$ dB gain for ${\mbox{IEEE1}}$.
For the ${\mbox{KATA2}}$ model we observe in Fig. \ref{fig:SNR_Out_Compare_Cycsta} an input SNR gain of $2.4$ dB compared to ${\mbox{Rx}}_3$  at ${\mbox{SNR}}_{in} \leq 0$ dB, which decreases to a $1.2$ dB gain at ${\mbox{SNR}}_{in} = 4$ dB.
However, when the impulsive noise component is very short, as in ${\mbox{KATA1}}$ (only $25$ microseconds impulse width) ${\mbox{Rx}}_4$ achieves relatively modest ${\mbox{SNR}}_{in}$ gains in the TA-MSE of about $1.2-0.35$ dB compared ${\mbox{Rx}}_3$ (smaller gains at higher values of ${\mbox{SNR}}_{in}$).
Observe that in all cases, as the direct recovery receiver (${\mbox{Rx}}_3$) exploits only the cyclostationary characteristics of the information signal, its performance improvement over the stationary Wiener filter (${\mbox{Rx}}_2$) is the same for both noise models at all SNRs. {\em The benefits of noise cancellation are thus clearly observed}.

\begin{figure}
\centering
\scalebox{0.5}{\includegraphics[clip=true,viewport=0.75in 0.2in 7.6in 5.6in]{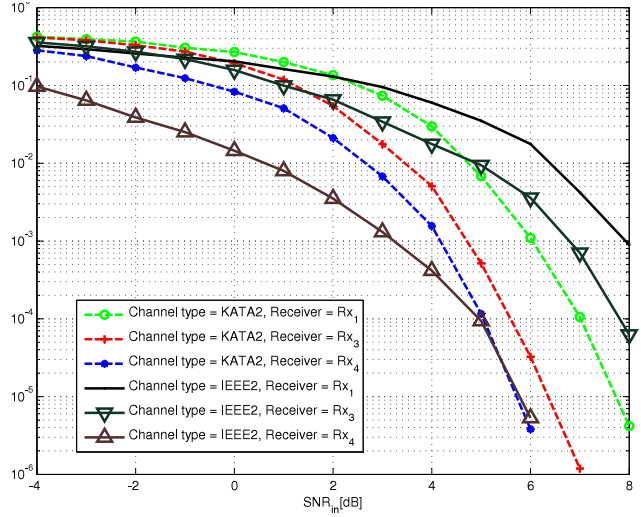}}
\vspace{-0.5cm}
\caption{BER comparison for different ACGN models.}
\label{fig:BER_Compare_Cycsta}
\vspace{-1.0cm}
\end{figure}

{ \bf{\em{3) BER Improvements in Coded Transmission Due to Noise Cancellation}:}}
The substantial gains obtained by ${\mbox{Rx}}_4$ in terms of TA-MSE translate directly into gain in BER. To demonstrate this point the coded BER results at the output of the different receivers for the ACGN channel are depicted in Fig. \ref{fig:BER_Compare_Cycsta} for both noise models. To avoid cluttering we depict only the results with the ${\mbox{IEEE2}}$ and the ${\mbox{KATA2}}$ parameters.
Observe that for the ${\mbox{IEEE2}}$ model the new receiver with noise cancellation (${\mbox{Rx}}_4$) achieves an input SNR gain of $5$ dB compared to ${\mbox{Rx}}_3$ at output BER of $10^{-1}$. This gain decreases to $4.3$ dB at output BER of $10^{-2}$ and to $3.5$ dB at output BER of $10^{-3}$.
For the ${\mbox{KATA2}}$ model the corresponding input SNR gains of ${\mbox{Rx}}_4$ over ${\mbox{Rx}}_3$ are $1.65$ dB, $0.75$ dB, and $0.55$ dB, respectively.
For the ${\mbox{IEEE1}}$ model these gains are $1.9$ dB, $1.25$ dB, and $1.2$ dB, respectively,
and for the ${\mbox{KATA1}}$ model these gains are $0.3$ dB, $0.27$ dB, and $0.25$ dB, respectively.
It is emphasized that this BER improvement is only due to noise cancellation.

\begin{figure}
\centering
\mbox
{
	\subfigure[Rx scatter plot for a given constellation point - no filtering.]
	{
		\includegraphics[width=2.7in,clip=true, viewport=0.75in 0.2in 7.6in 5.6in]{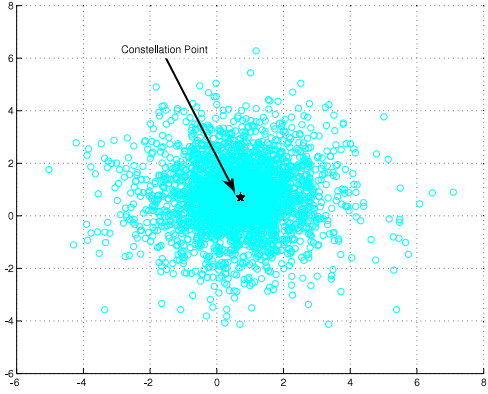}
		\label{fig:Bias_no_Filt}
	}	
	\quad
	\subfigure[Rx scatter plot for a given constellation point - after FRESH filtering with noise cancellation.]
	{
		\includegraphics[width=2.7in,clip=true, viewport=0.75in 0.2in 7.6in 5.6in]{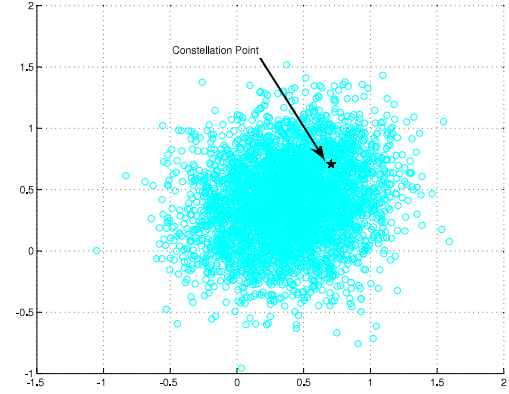}
		\label{fig:Bias_FRESH_Filt}
	}
}
\vspace{-0.35cm}
\caption{Scaling induced by filter analysis for the Katayama noise channel, ${\mbox{SNR}}_{in}=-4$ dB.}
\label{fig:Bias_Compare}
\vspace{-0.6cm} 
\end{figure}

\begin{figure}
\centering
\begin{minipage}{.45\linewidth}
  \includegraphics[width=3in,clip=true, viewport=0.75in 0.2in 7.6in 5.6in]{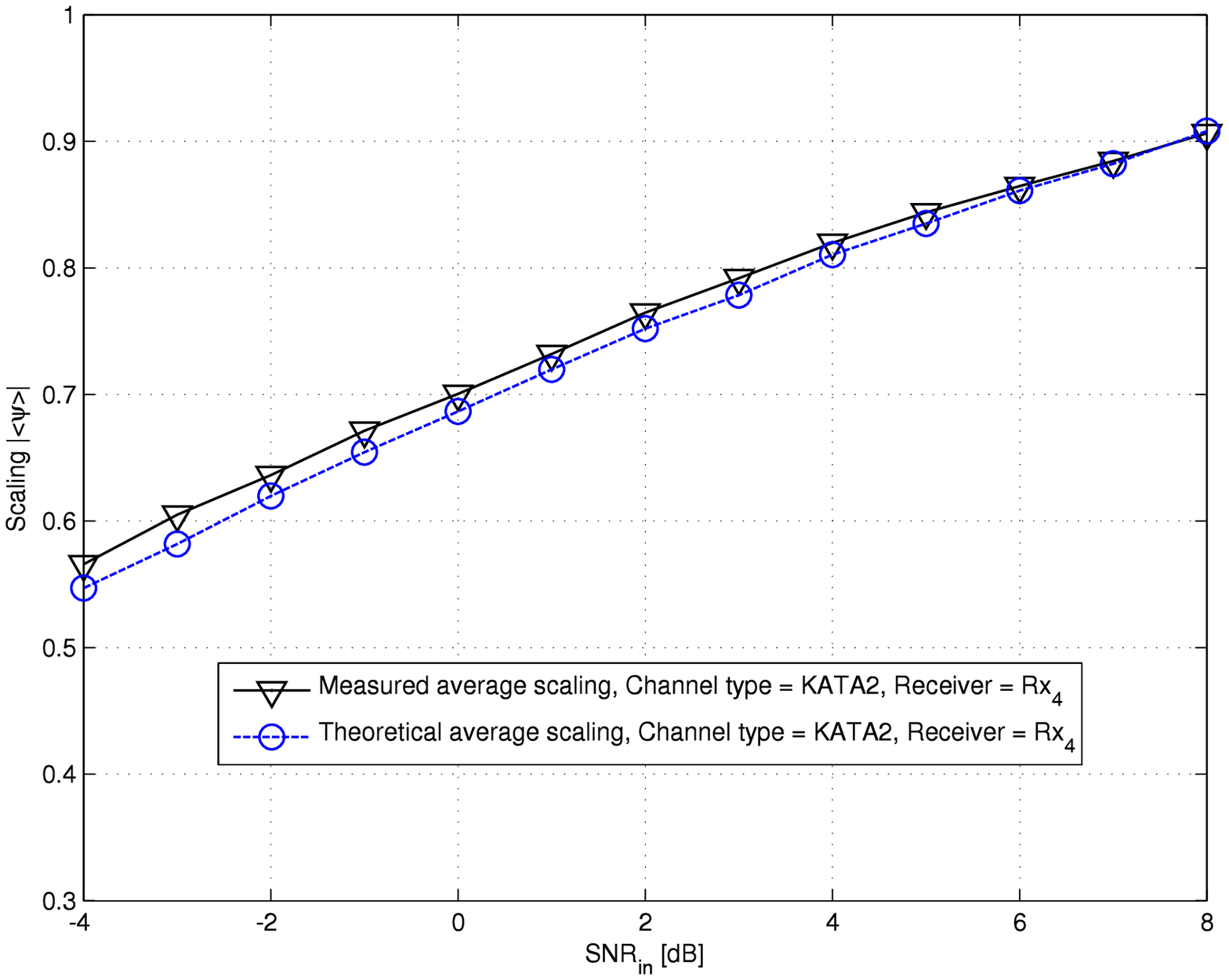}
  \vspace{-0.9cm}
  \caption{Comparison of constellation scaling at the output of ${\mbox{Rx}}_4$ obtained from simulation and from analysis.}
  \label{fig:Scale_Compare_Cycsta}
\end{minipage}
\quad
\begin{minipage}{.45\linewidth}
  \includegraphics[width=3in,clip=true, viewport=0.75in 0.2in 7.6in 5.6in]{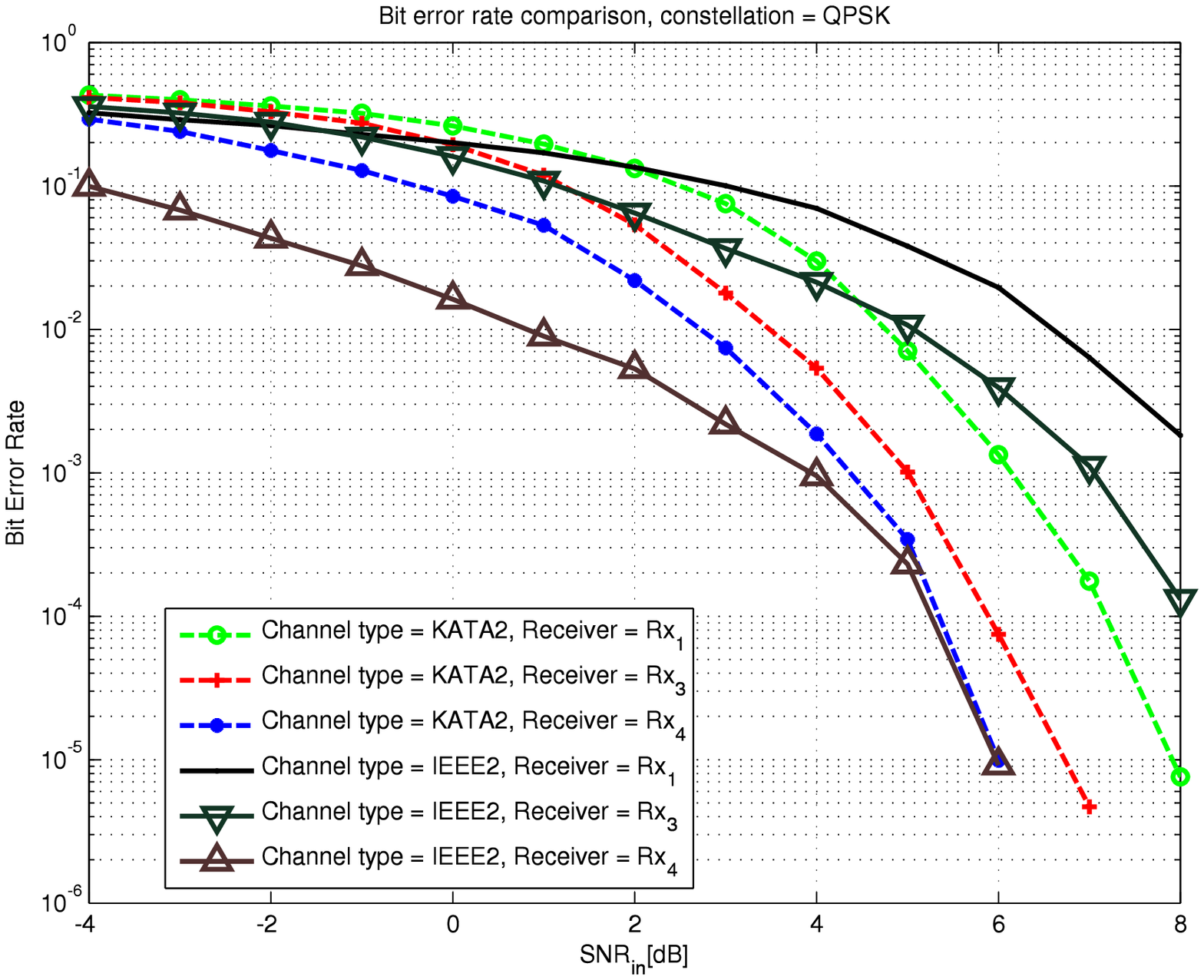}
  \vspace{-0.9cm}
  \caption{BER results for 4-tap multipath for different ACGN models.}
  \label{fig:BER_Multipath_CYCSTA}
\end{minipage}
\vspace{-1.0cm} 
\end{figure}

Note that while the coded BER gain for our proposed receiver (${\mbox{Rx}}_4$) compared to the direct recovery receiver (${\mbox{Rx}}_3$) corresponds to the TA-MSE improvement depicted in Figs. \ref{fig:SNR_Out_Compare_LPTV} and \ref{fig:SNR_Out_Compare_Cycsta}, there is a difference between the ${\mbox{SNR}}_{in}$ gain for the coded BER and the ${\mbox{SNR}}_{in}$ gain for the TA-MSE when ${\mbox{Rx}}_4$ is compared to ${\mbox{Rx}}_1$:
The proposed receiver ${\mbox{Rx}}_4$ achieves coded BER of $10^{-2}$ for the Katayama noise model ${\mbox{KATA2}}$  at ${\mbox{SNR}}_{in}$ of $2.65$ dB, while the receiver with no filtering (${\mbox{Rx}}_1$) achieves the same coded BER for ${\mbox{SNR}}_{in}$ of $4.75$ dB, i.e., the ${\mbox{SNR}}_{in}$ gain for coded BER of $10^{-2}$ is $2.1$ dB. However, the TA-MSE at the output of ${\mbox{Rx}}_4$, for ${\mbox{SNR}}_{in}$ of $2.65$ dB is $-21.3$ dB, which correspond to an ${\mbox{SNR}}_{in}$ gain of $3.4$ dB over ${\mbox{Rx}}_1$.
It follows that a $3.4$ dB gain in TA-MSE translates into a $2.1$ dB gain in coded BER.
The reason is the scaling effect discussed in Subsection \ref{subsec:Scaling},
which is demonstrated in Fig. \ref{fig:Bias_Compare}. Fig. \ref{fig:Bias_no_Filt} depicts the received subcarrier values corresponding to a specific QPSK constellation symbol for transmission over the ${\mbox{KATA2}}$ noise channel model at ${\mbox{SNR}}_{in}=-4$ dB without any filtering at the receiver (${\mbox{Rx}}_1$), while Fig. \ref{fig:Bias_FRESH_Filt} depicts the received values for the same QPSK symbol at the output of the newly proposed receiver, ${\mbox{Rx}}_4$, for the same scenario of Fig. \ref{fig:Bias_no_Filt}. It can be seen that for ${\mbox{Rx}}_4$, although the received values are closer to the transmitted constellation symbol as compared to the values without filtering, a scaling effect is induced, resulting in a smaller uncoded BER improvement compared to the improvement in TA-MSE.
Fig. \ref{fig:Scale_Compare_Cycsta} compares between the values of the scaling obtained from the simulation and from the analytical expression \eqref{eqn:scale_14} for ${\mbox{Rx}}_4$ with the ${\mbox{KATA2}}$ model parameters. The measured average scaling is obtained by averaging the scaling of the recovered signal at the output of $h_2[n]$, that is, evaluating $\left\langle\frac{y[n]}{d[n]} \right\rangle$. Observe that the simulated results agree with the analytical derivation, especially at high SNR, confirming the validity of the scaling analysis in Subsection \ref{subsec:Scaling}.

{ \bf{\em{4) Applicability of the New Receiver with Noise Cancellation (${\mbox{Rx}}_4$) to ISI Channels}:}}
Lastly, we examine the application of the receiver with noise cancellation (${\mbox{Rx}}_4$) to ISI channels. The simulation was carried out for a 4-tap multipath channel with an exponentially decaying attenuation profile, as in~\cite{Chen:11}, with tap values $[1, 0.1, 0.01, 0.001]$, applied prior to the addition of the noise. Fig. \ref{fig:BER_Multipath_CYCSTA} depicts the results for the ACGN for both models with the ${\mbox{IEEE2}}$ and the ${\mbox{KATA2}}$ parameter sets. Note that the results and the corresponding SNR gains are similar to the results obtained for the channel without ISI, since, as described in Subsection \ref{subsec:Multipath}, the multipath channel effect is accounted for by designing the FRESH filter to recover the OFDM signal {\em after} convolution with the channel. {\em This shows that the new receiver is very beneficial for multipath channels and not only for memoryless channels}.

{ \bf{\em{5) Sensitivity of the New Receiver with Noise Cancellation (${\mbox{Rx}}_4$) to Cyclic Frequency Errors}:}}
The frequency shift filtering algorithm proposed in the present paper requires the receiver to know the exact period of the noise in order to compute the cyclic frequencies.
    Clearly, as stated in \cite{Ojeda:11}, error in the knowledge of the cyclic frequencies severely damages the performance of the FRESH filter. This is because when the error is high enough, the FRESH filter input operates on shifted versions of the signal which do not have cyclic correlation, thus, all branches of the FRESH filter become useless, except the branch which performs no frequency shift ($\alpha_{k_0} = 0$).
    
     However, this is not an issue in narrowband PLC as the period of the noise is known to be half the AC cycle (see \cite{Pavlidou:03}, \cite{Hooijen:98}, \cite{Zimmermann:02a}, \cite{Zimmermann:02b}, \cite{Katayama:06}, \cite{Nassar:12}). Note also that the period of the OFDM signal is known by design.

\begin{figure}
\centering
\mbox
{
	\subfigure[Cyclic frequency error sensitivity analysis  - IEEE1 parameters.]
	{
		\includegraphics[width=2.7in,clip=true, viewport=0.75in 0.2in 7.6in 5.6in]{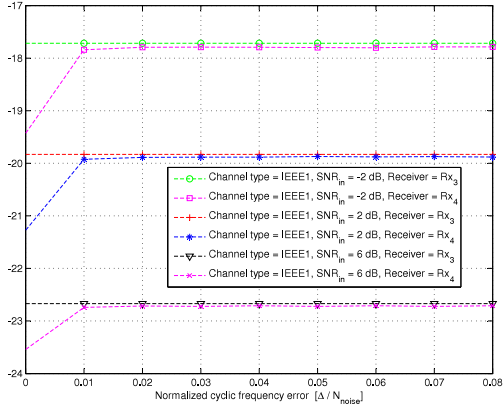}
		\label{fig:Sensitivity_IEEE1}
	}	
	\quad
	\subfigure[Cyclic frequency error sensitivity analysis  - IEEE2 parameters.]
	{
		\includegraphics[width=2.7in,clip=true, viewport=0.75in 0.2in 7.6in 5.6in]{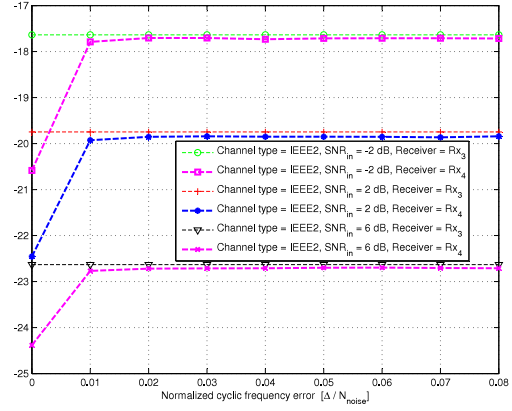}
		\label{fig:Sensitivity_IEEE2}
	}
}
\vspace{-0.25cm}
\caption{TA-MSE comparison in the presence of an error in the cyclic frequency of the noise - ACGN model of \cite{Nassar:12}.}
\label{fig:Sensitivity_IEEE}
\vspace{-0.4cm} 
\end{figure}

\begin{figure}
\centering
\mbox
{
	\subfigure[Cyclic frequency error sensitivity analysis  - KATA1 parameters.]
	{
		\includegraphics[width=2.7in,clip=true, viewport=0.75in 0.2in 7.6in 5.6in]{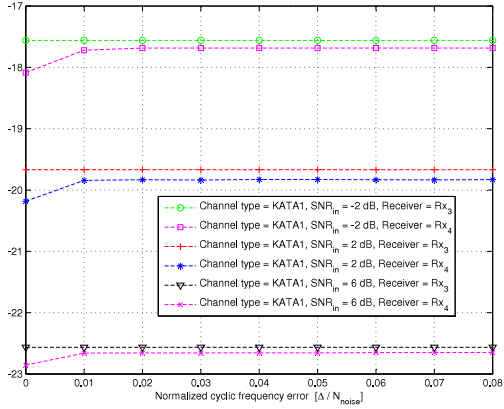}
		\label{fig:Sensitivity_KATA1}
	}	
	\quad
	\subfigure[Cyclic frequency error sensitivity analysis  - KATA2 parameters.]
	{
		\includegraphics[width=2.7in,clip=true, viewport=0.75in 0.2in 7.6in 5.6in]{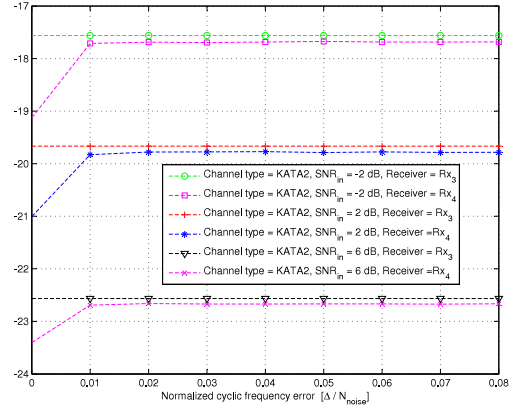}
		\label{fig:Sensitivity_KATA2}
	}
}
\vspace{-0.25cm}
\caption{TA-MSE comparison in the presence of an error in the cyclic frequency of the noise - ACGN model of \cite{Katayama:06}.}
\label{fig:Sensitivity_KATA}
\vspace{-0.5cm} 
\end{figure}

    Nonetheless, in order to analyze the performance degradation in case there actually is an error in the knowledge of the cyclic frequencies of the noise, we carried out simulations for all four ACGN noise parameter sets, in the presence of an error in the cyclic frequencies of the noise used by the receiver (we denote the error with $\Delta$). The simulations compared the TA-MSE of the new receiver ${\mbox{Rx}}_4$ and the FRESH filter without noise cancellation of \cite{Chen:11} (${\mbox{Rx}}_3$) for ${\mbox{SNR}}_{in} = \{-2, 2, 6\}$ dB. The results are depicted in Fig. \ref{fig:Sensitivity_IEEE} of this letter for the IEEE LPTV noise model of \cite{Nassar:12}, and in Fig. \ref{fig:Sensitivity_KATA} of this letter for the Katayama noise model of \cite{Katayama:06}. We observe from the figures that, as expected, when the error in the cyclic frequencies of the noise is large enough, the noise cancellation FRESH filter $h_1[n]$ becomes ineffective, and the TA-MSE performance of ${\mbox{Rx}}_4$ converges to that of ${\mbox{Rx}}_3$. It is important to note that the cyclic frequency error does not degrade the performance of ${\mbox{Rx}}_4$ compared to ${\mbox{Rx}}_3$.

\vspace{-0.5cm}
\subsection{Simulation Study of the Adaptive Optimal Receiver}
\label{sec:Adaptive_Sim}
\vspace{-0.25cm}
We now turn to evaluate the performance of the adaptive implementation developed in Section \ref{sec:Adaptive}.
Three receivers are simulated - ${\mbox{Rx}}_1$ (no filtering), the adaptive version of ${\mbox{Rx}}_3$, and the adaptive version of ${\mbox{Rx}}_4$. Here, two main aspects were tested: the convergence of the adaptive filter to the optimal solution and the robustness to the noise model.

\begin{figure}
\centering
\begin{minipage}{.45\linewidth}
  \includegraphics[width=3in,clip=true, viewport=0.75in 0.2in 7.6in 5.6in]{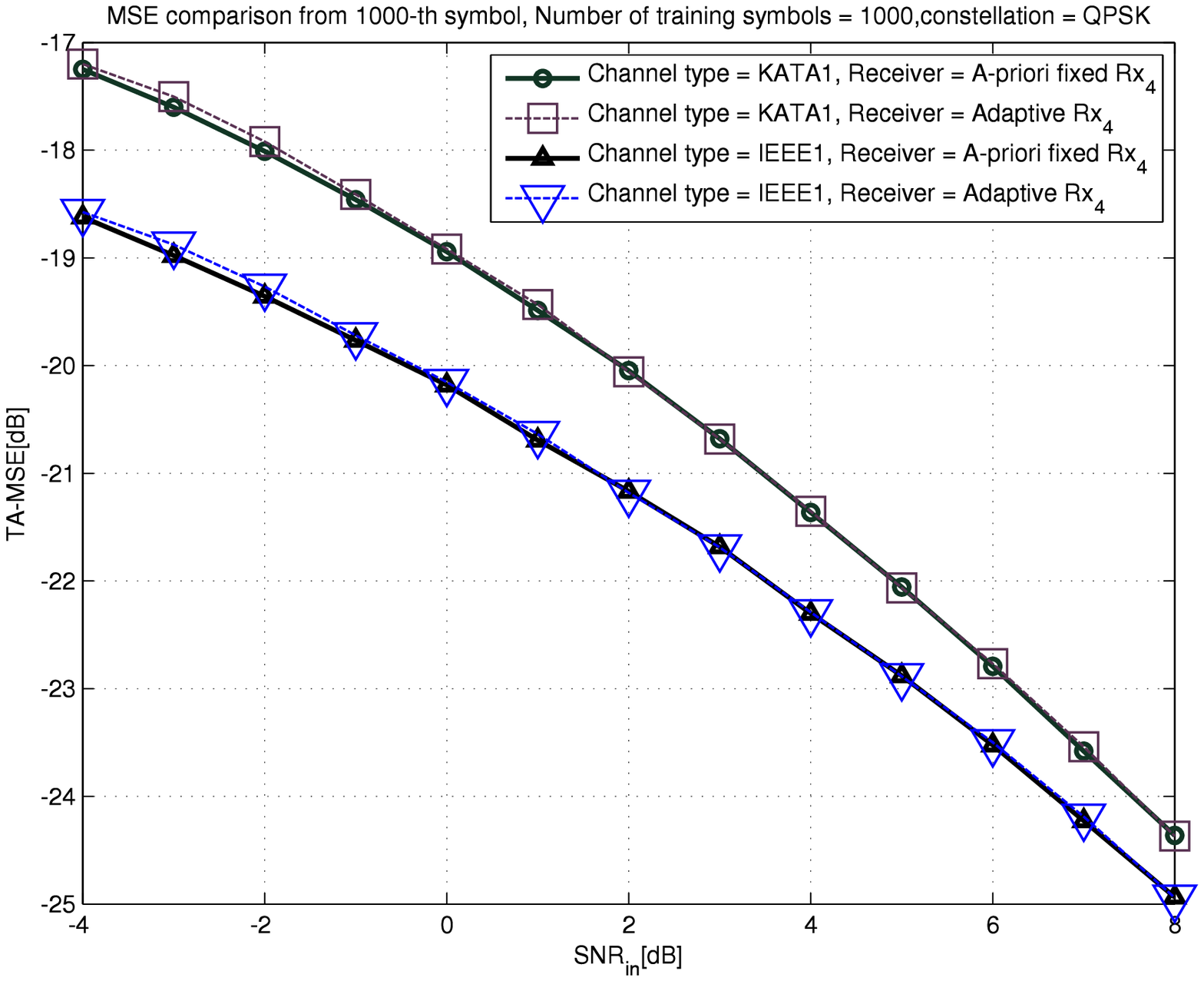}
  \vspace{-0.8cm}
  \caption{Convergence of the TA-MSE of the adaptive ${\mbox{Rx}}_4$ with training to the TA-MSE of the optimal ${\mbox{Rx}}_4$.}
  \label{fig:SNR_PerfectFRESH}
\end{minipage}
\quad
\begin{minipage}{.45\linewidth}
  \includegraphics[width=3in,clip=true, viewport=0.75in 0.2in 7.6in 5.6in]{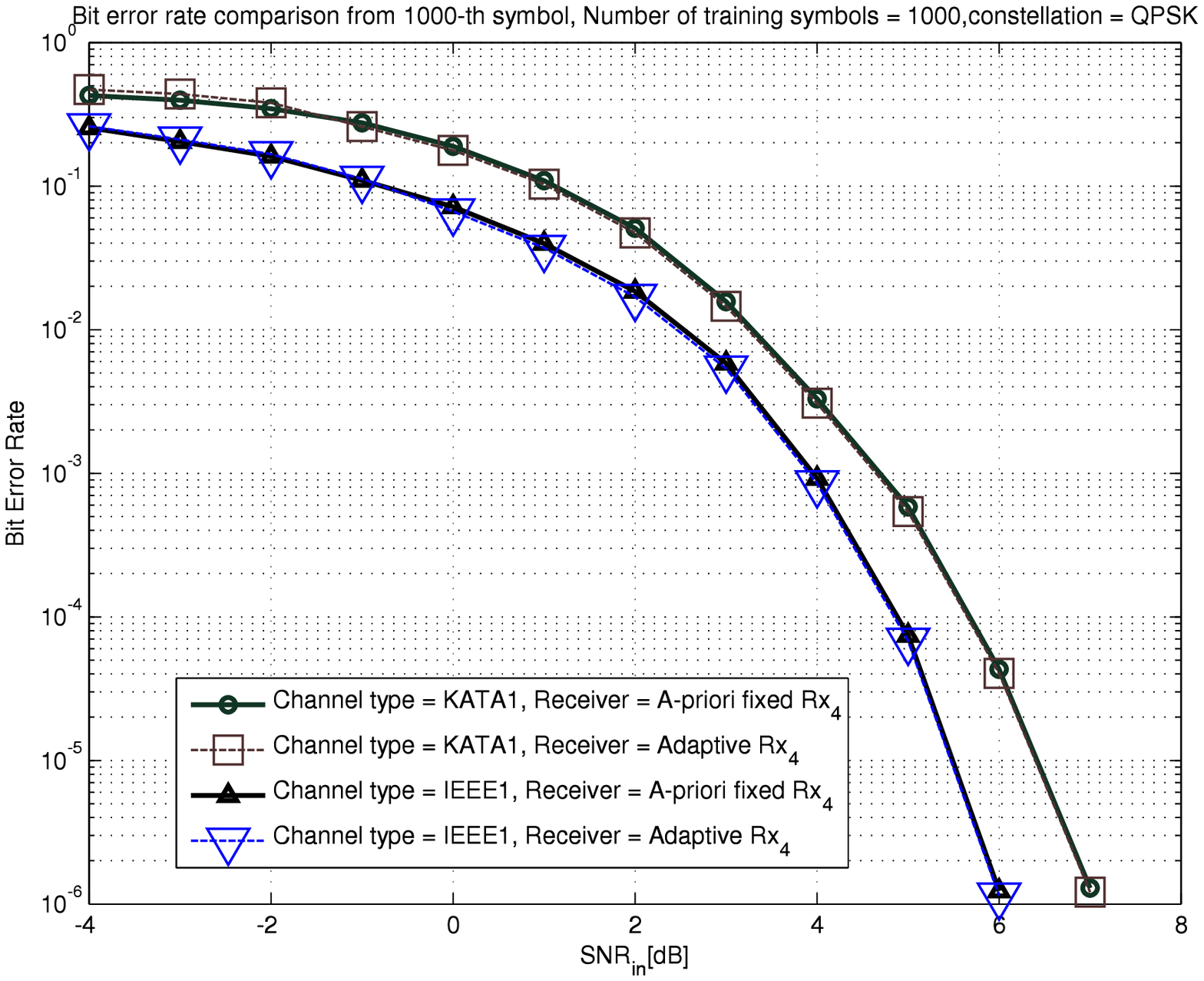}
  \vspace{-0.8cm}
  \caption{Convergence of the BER of the adaptive ${\mbox{Rx}}_4$ with training to the BER of the optimal ${\mbox{Rx}}_4$.}
  \label{fig:BER_PerfectFRESH}
\end{minipage}
\vspace{-1.0cm} 
\end{figure}

{ \bf{\em{1) Convergence of the Adaptive Filter to the Optimal Solution}:}}
First, we examined the performance of the adaptive FRESH filter for an error-free reference signal. This models communications with sufficiently strong error-correction codes.
Fig. \ref{fig:SNR_PerfectFRESH} compares the TA-MSE of the adaptive ${\mbox{Rx}}_4$ with an ideal reference signal and the TA-MSE of the optimal ${\mbox{Rx}}_4$ derived in Section \ref{sec:Model}, for two ACGN models: the LPTV model~\cite{Nassar:12} with parameters set ${\mbox{IEEE1}}$ and the Katayama model~\cite{Katayama:06} with parameters set ${\mbox{KATA1}}$. BER comparison is depicted in Fig. \ref{fig:BER_PerfectFRESH}.
As expected, the performance of the adaptive ${\mbox{Rx}}_4$ with an error-free reference signal converges to that of the optimal ${\mbox{Rx}}_4$ for both models.

\begin{figure}
\centering
\begin{minipage}{.45\linewidth}
  \includegraphics[width=3in,clip=true, viewport=0.75in 0.2in 7.6in 5.6in]{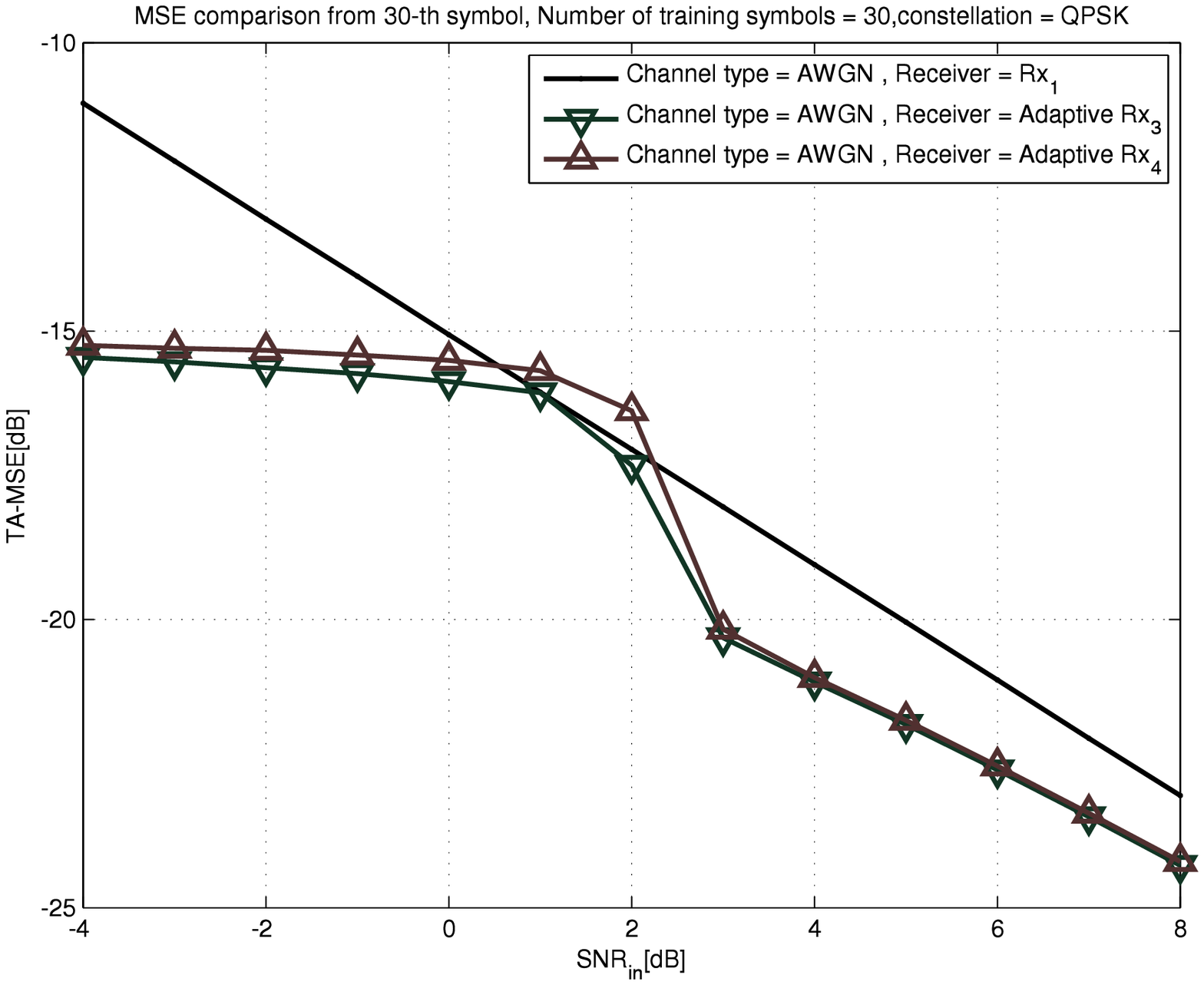}
  \vspace{-0.8cm}
  \caption{Testing robustness of the new algorithm to the noise model: TA-MSE of the decision directed adaptive ${\mbox{Rx}}_3$ and ${\mbox{Rx}}_4$ for the AWGN channel.}
  \label{fig:MSE_Adaptive_AWGN}
\end{minipage}
\quad
\begin{minipage}{.45\linewidth}
  \includegraphics[width=3in,clip=true, viewport=0.75in 0.2in 7.6in 5.6in]{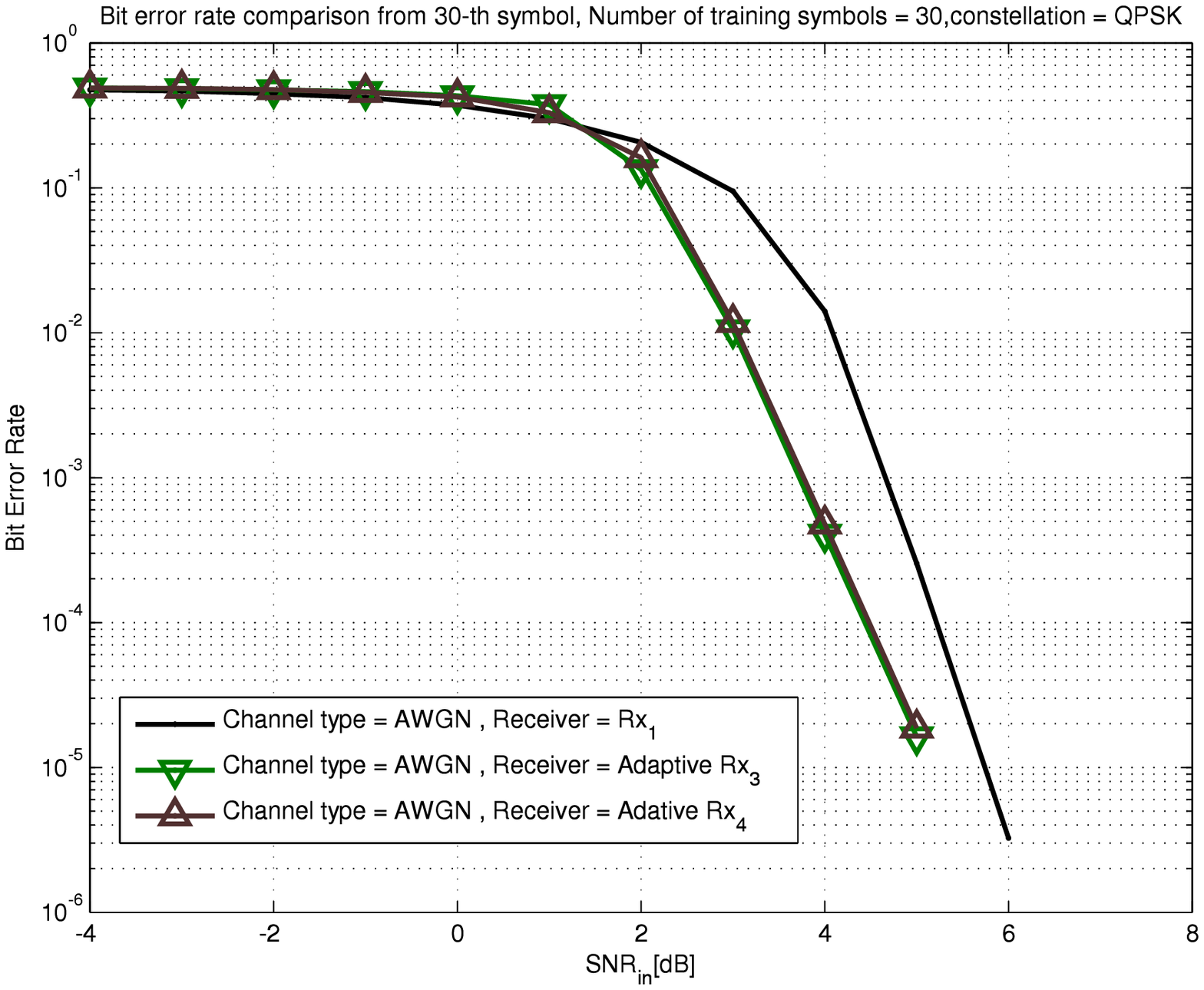}
  \vspace{-0.8cm}
  \caption{Testing robustness of the new algorithm to the noise model: BER of the decision directed adaptive ${\mbox{Rx}}_3$ and ${\mbox{Rx}}_4$ for the AWGN channel.}
  \label{fig:BER_Adaptive_AWGN}
\end{minipage}
\vspace{-1.0cm} 
\end{figure}

{ \bf{\em{2) Verifying Robustness of the Adaptive Algorithm to Noise Model and Corresponding Performance}:}}
We next verified that the adaptive implementation operates well also in AWGN. 
The simulation results are presented in Fig. \ref{fig:MSE_Adaptive_AWGN} and in Fig. \ref{fig:BER_Adaptive_AWGN}. Observe that for ${\mbox{SNR}}_{in}\geq 3$ dB the decision-directed adaptive ${\mbox{Rx}}_4$ obtains the same TA-MSE performance as the optimal ${\mbox{Rx}}_4$ shown in Fig. \ref{fig:SNR_Out_Compare_AWGN}.
Also observe that, as expected, the results for the AWGN channel show that applying the noise cancellation filter (${\mbox{Rx}}_4$) does not improve upon the single FRESH filter of ${\mbox{Rx}}_3$, and in fact, for ${\mbox{SNR}}_{in}\geq 3$ dB the performance of the adaptive ${\mbox{Rx}}_4$ is roughly equal to that of the adaptive ${\mbox{Rx}}_3$. This is because the noise exhibits no cyclic redundancy. The unfavorable performance observed for ${\mbox{SNR}}_{in}\leq 3$ dB is due to the fact that the BER it too high for the adaptive algorithm to converge, thus, if it is desired to work at lower SNRs, stronger coding is needed.

\begin{figure}
\centering
\begin{minipage}{.45\linewidth}
  \includegraphics[width=3in,clip=true, viewport=0.75in 0.2in 7.6in 5.6in]{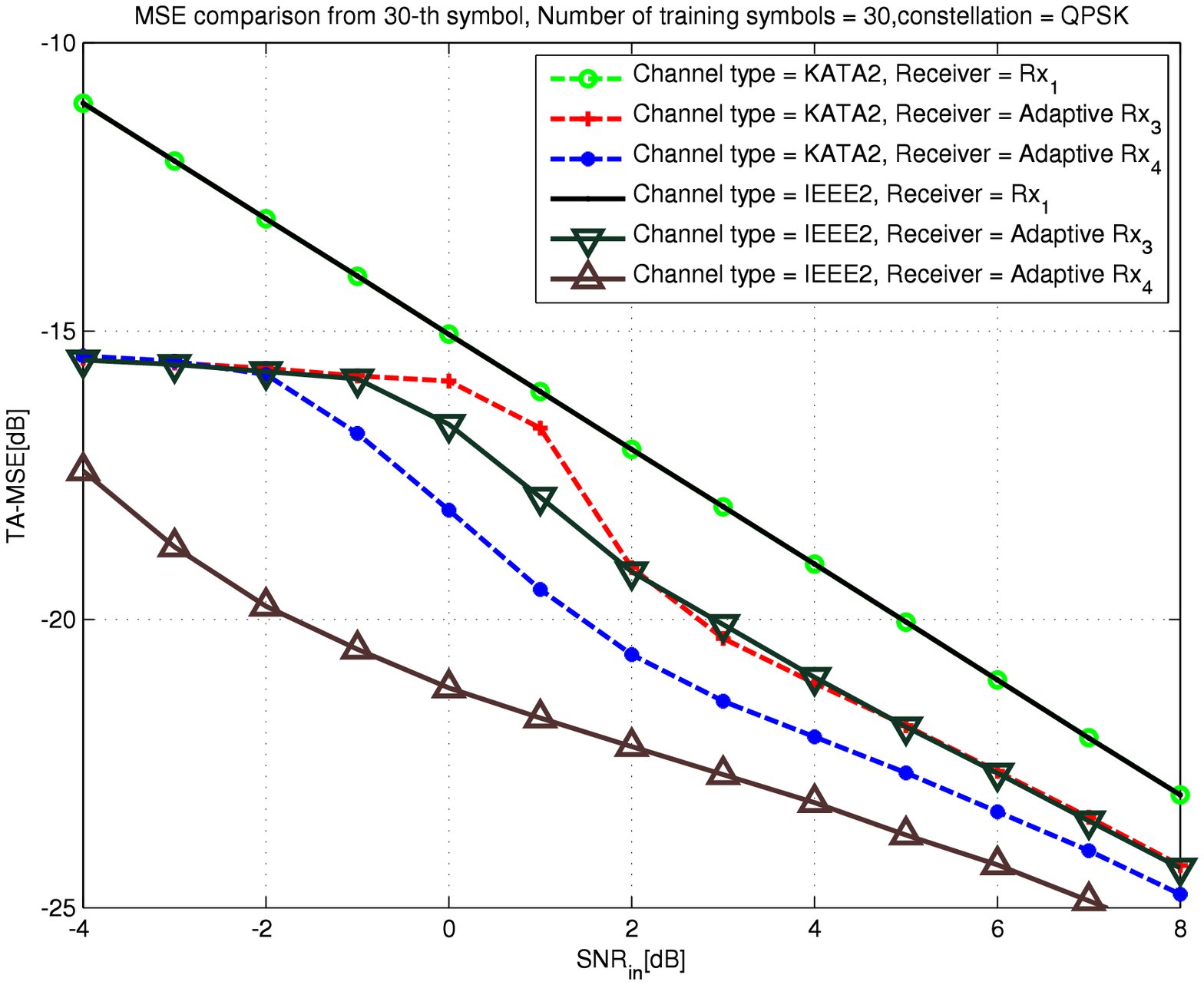}
  \vspace{-0.8cm}
  \caption{TA-MSE comparison for the adaptive ${\mbox{Rx}}_3$ and ${\mbox{Rx}}_4$ for ACGN channels.}
  \label{fig:MSE_Adaptive_CYCSTA}
\end{minipage}
\quad
\begin{minipage}{.45\linewidth}
  \includegraphics[width=3in,clip=true, viewport=0.75in 0.2in 7.6in 5.6in]{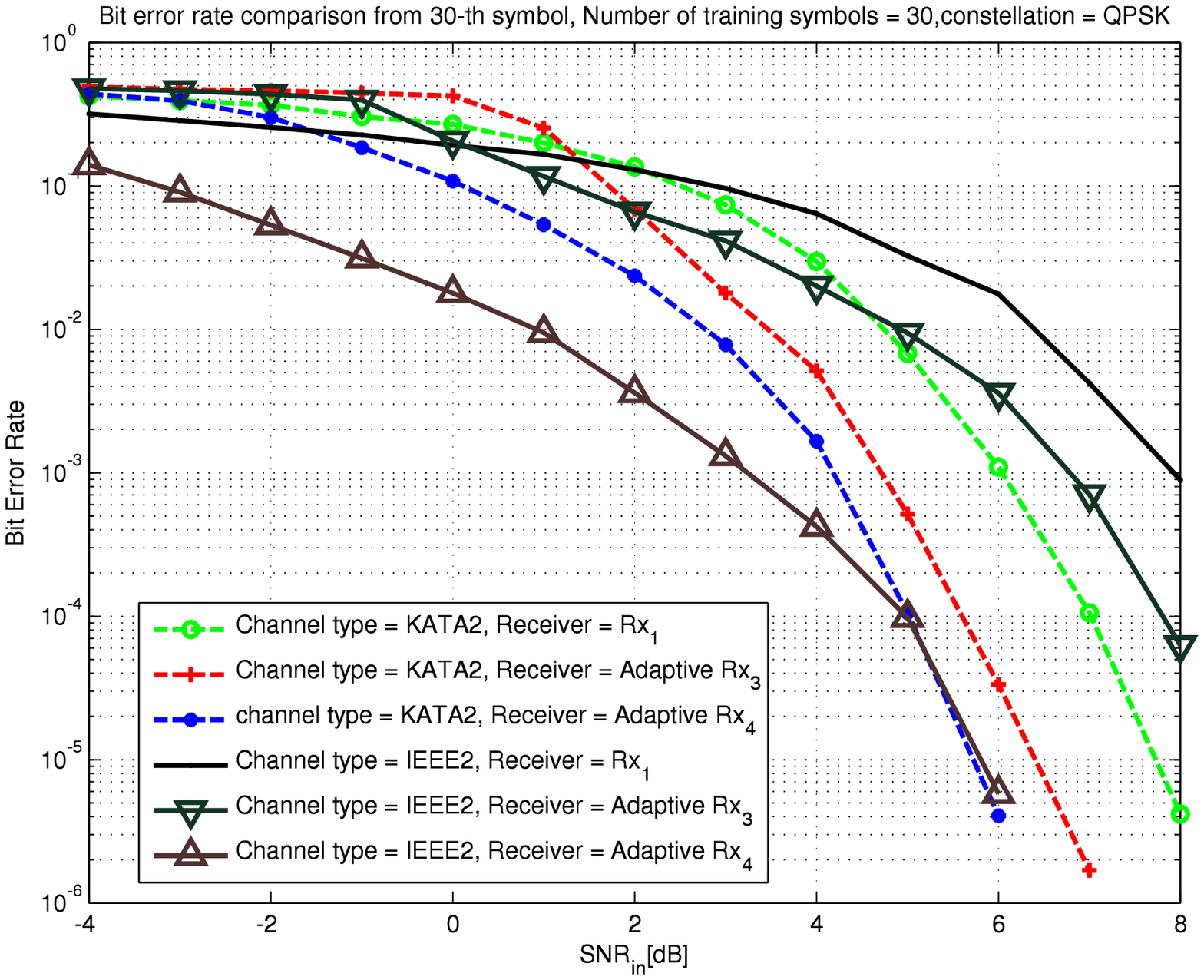}
  \vspace{-0.8cm}
  \caption{BER comparison for the adaptive ${\mbox{Rx}}_3$ and ${\mbox{Rx}}_4$ for ACGN channels.}
  \label{fig:BER_Adaptive_CYCSTA}
\end{minipage}
\vspace{-1.0cm} 
\end{figure}

Lastly, the TA-MSE and the BER performance were evaluated for two sets of ACGN models: The LPTV model \cite{Nassar:12} with parameters ${\mbox{IEEE2}}$ and the Katayama model \cite{Katayama:06} with parameters ${\mbox{KATA2}}$. The TA-MSE and the BER results are depicted in Fig. \ref{fig:MSE_Adaptive_CYCSTA} and Fig. \ref{fig:BER_Adaptive_CYCSTA}, respectively.
Observe that for both models the adaptive ${\mbox{Rx}}_4$ is beneficial for all ${\mbox{SNR}}_{in}$ values.
For the ${\mbox{IEEE2}}$ model the adaptive ${\mbox{Rx}}_4$ converges to the optimal ${\mbox{Rx}}_4$ for ${\mbox{SNR}}_{in}\geq 0$ dB which corresponds to BER values lower than $2\cdot10^{-2}$, while the adaptive ${\mbox{Rx}}_3$ converges to the optimal ${\mbox{Rx}}_3$ for ${\mbox{SNR}}_{in}\geq 3$ dB which corresponds to BER values lower than $4\cdot10^{-2}$. This is due to the effect of the output BER on the decision-based adaptive implementation.
For the ${\mbox{KATA2}}$ model the adaptive ${\mbox{Rx}}_4$ converges to the optimal ${\mbox{Rx}}_4$ for ${\mbox{SNR}}_{in}\geq 2.25$ dB while the adaptive ${\mbox{Rx}}_3$ converges to the optimal ${\mbox{Rx}}_3$ for ${\mbox{SNR}}_{in}\geq 3$ dB, both correspond to BER values lower than $2\cdot10^{-2}$.
Note that \cite{IEEE:13} defines the appropriate working region for narrowband PLC 
as the situation in which packet error rate (PER) for a packet consisting of 100 octets is less than $0.1$. As this PER corresponds to BER of less than $1.3\cdot10^{-4}$, it follows that the adaptive ${\mbox{Rx}}_4$ converges to the optimal ${\mbox{Rx}}_4$ within the appropriate working region.
It is therefore concluded that the substantial gains obtained by the optimal ${\mbox{Rx}}_4$ can be obtained also with the adaptive implementation.

\vspace{-0.25cm}
\section{Conclusions}
\label{sec:Conclusions}
\vspace{-0.25cm}
In this paper, a new receiver designed for exploiting the cyclostationary characteristics of the OFDM information signal as well as {\em those of the narrowband PLC channel noise} is proposed. The novel aspect of the work is the insight that {\em noise estimation} in cyclostationary noise channels is {\em beneficial}, contrary to the widely used AWGN channels. It was shown that at low SNRs, which characterize the narrowband PLC channel, a substantial performance improvement can be obtained by signal recovery combined with noise cancellation via FRESH filtering, compared to previous approaches which focused on estimating only the information signal. This gain was demonstrated for different cyclostationary noise models and in particular for the noise models specified in the IEEE standard~\cite{IEEE:13}. It was also shown that with an appropriate design, the proposed model can be applied also to ISI channels.
We then presented an adaptive implementation of the receiver, and identified the BER range in which this implementation is beneficial.
Future work will focus on adopting the cyclostationary signal processing schemes to the MIMO narrowband PLC channel.

\end{document}